\documentclass[11pt]{article} 
\usepackage{fullpage}
\usepackage{graphicx}
\usepackage{upref}
\usepackage{enumerate}
\usepackage{latexsym}
\usepackage{pdfpages}
\usepackage[bookmarks]{hyperref}
\usepackage{color,graphics}
\usepackage{comment} 
\usepackage{caption}
\usepackage{subcaption}
\usepackage{wrapfig}
\usepackage{braket}

\usepackage{amssymb}
\usepackage{amsmath}
\usepackage{amsthm}
\usepackage{mathrsfs}
\usepackage{mathtools}

\usepackage{epsfig,graphicx}
\usepackage{algorithmic}
\usepackage[ruled,linesnumbered]{algorithm2e}
\usepackage{amsfonts}
\usepackage{tikz} 
\usepackage{varioref}
\usepackage[ansinew]{inputenc}
\usepackage{qcircuit}
\usepackage{authblk}
\usepackage{multirow}

\theoremstyle{plain}

\theoremstyle{plain}
\newtheorem{lemma}{Lemma}[section]
\theoremstyle{plain}

\theoremstyle{plain}

\theoremstyle{plain}

\theoremstyle{plain}

\theoremstyle{definition}

\newtheorem{definition}{Definition}[section]
\theoremstyle{definition}
\newtheorem{fact}{Fact}[section]

\theoremstyle{remark}

\theoremstyle{definition}

\AtBeginDocument{}

\DeclareMathOperator{\real}{\mathbb{R}}
\DeclareMathOperator{\nat}{\mathbb{N}}

\newcommand{\intg}{\mathbb{Z}}

\newcommand{\fld}{\mathbb{F}}

\newcommand{\id}{\mathbb{I}}

\newcommand{\cliff}{\mathcal{C}}
\newcommand{\pauli}{\mathcal{P}}

\newcommand{\unit}{\mathcal{U}}

\newcommand{\X}{\text{X}}
\newcommand{\Y}{\text{Y}}
\newcommand{\Z}{\text{Z}}
\newcommand{\had}{\text{H}}
\newcommand{\T}{\text{T}}
\newcommand{\CNOT}{\text{CNOT}}
\newcommand{\phase}{\text{S}}

\newcommand{\vect}[1]{\textbf{#1}}
\newcommand{\ckt}{\mathcal{C}}
\newcommand{\phpoly}{\mathcal{P}}
\newcommand{\Span}{\text{span}}
\newcommand{\hset}{\mathcal{H}}
\newcommand{\stack}{\mathcal{K}}
\newcommand{\tset}{\mathcal{S}}

\begin{document}

\title{Reducing the CNOT count for Clifford+T circuits on NISQ architectures}

\author[1,2]{Vlad Gheorghiu \thanks{vlad.gheorghiu@uwaterloo.ca}}
\author[3]{Jiaxin Huang \thanks{j2huang@uwaterloo.ca}}
\author[1,4]{Sarah Meng Li \thanks{sarah.li@dal.ca}}
\author[1,2,4,5]{Michele Mosca \thanks{michele.mosca@uwaterloo.ca}}
\author[1,4]{Priyanka Mukhopadhyay \thanks{Corresponding author : mukhopadhyay.priyanka@gmail.com, p3mukhop@uwaterloo.ca}}

\affil[1]{Institute for Quantum Computing, University of Waterloo, Waterloo ON, Canada}
\affil[2]{softwareQ Inc., Kitchener ON, Canada}
\affil[3]{Faculty of Mathematics, University of Waterloo, Canada}
\affil[4]{Department of Combinatorics and Optimization, University of Waterloo, Waterloo ON, Canada}
\affil[5]{Perimeter Institute for Theoretical Physics, Waterloo, Waterloo ON, Canada}

\maketitle

\begin{abstract}
While mapping a quantum circuit to the physical layer one has to consider the numerous constraints imposed by the underlying hardware architecture. Connectivity of the physical qubits is one such constraint that restricts two-qubit operations, such as CNOT, to ``connected'' qubits. SWAP gates can be used to place the logical qubits on admissible physical qubits, but they entail a significant increase in CNOT-count.
In this paper we consider the problem of reducing the CNOT-count in Clifford+T circuits on connectivity constrained architectures, like noisy intermediate-scale quantum (NISQ) computing devices. We ``slice'' the circuit at the position of Hadamard gates and ``build'' the intermediate $\{\CNOT,\T\}$ sub-circuits using Steiner trees, significantly improving on previous methods.
We compared the performance of our algorithms while mapping different benchmark and random circuits to some well-known architectures such as 9-qubit square grid, 16-qubit square grid, Rigetti 16-qubit Aspen, 16-qubit IBM QX5 and 20-qubit IBM Tokyo. Our methods give less CNOT-count compared to Qiskit and TKET transpiler as well as using SWAP gates. 
Assuming most of the errors in a NISQ circuit implementation are due to CNOT errors, then our method would allow circuits with few times more CNOT gates be reliably implemented than the previous methods would permit.
\end{abstract}



\section{Introduction}\label{sec:Intro}

Quantum computing is a computational paradigm which is predicted to provide significant speedups for problems including, but not limited to, large number factorization~\cite{1999_S}, simulation of quantum systems~\cite{1982_F} and unstructured search~\cite{1997_G}, all of which are believed to be intractable or significantly slower on a classical computer. Somewhat similar to its classical counterpart, a quantum circuit consisting of elementary unitary operations remains the most popular model for quantum computation. Thus we need efficient quantum compilers that map a high-level algorithm into a lower-level form, that is, a quantum circuit consisting of quantum gates that are admissible by the hardware constraints.

At present we do not have large-scale quantum computers. Rather the devices available today are referred to as noisy intermediate-scale quantum (NISQ) computers \cite{2018_P}. The current technologies that realize these devices, such as superconducting quantum circuits \cite{2018_ROTetal, 2017_VPKetal} and ion traps \cite{2012_BSKetal, 2016_BHLSL, 2016_GTLetal, 2006_HOSetal}, impose certain connectivity constraints by which two-qubit operations are possible only among certain pairs of physical qubits. Naively we can insert SWAP operators to move a pair of logical\footnote{
In NISQ systems, a ``logical qubit'' typically corresponds to a single individual qubit, in contrast to fault-tolerant quantum computation where a logical qubit is encoded in many physical qubits. However, the correspondence between the logical qubits and physical qubits on a NISQ computer can change throughout the computation.} 
qubits to physical positions admissible for two-qubit operations. However, this increases the number of two-qubit operations, each of which again introduces non-negligible noise. Hence, it is important to optimize the number of two-qubit operators while respecting the connectivity constraints. 

In this paper we consider the problem of re-synthesizing a circuit over the universal fault-tolerant Clifford+T gate set. We designed and implemented an algorithm that reduces the number of CNOT gates required to meet the connectivity constraints imposed by the physical hardware architectures. The connectivity constraints are represented in the form of a graph $G$ (called \textbf{connectivity graph}) in which the vertices represent (physical) qubits and a two-qubit operation can be applied if and only if the corresponding vertices are connected by an edge in $G$. We assume without loss of generality that 
the desired circuit is connected (that is, we can't break it up into non-interacting collections of qubits).

The Clifford+T gate set is one of the most studied fault-tolerant universal gate set used to realize a quantum operator. 
We consider the following gates in this set: $\scriptsize{\mathcal{G}_{univ}=\{\CNOT,\had,\T,\T^{\dagger},\phase,\phase^{\dagger},\X,\Y,\Z\}}$, among which CNOT  is the only multi-qubit operator. 
If the shortest path length between vertices corresponding to $c$ and $t$ in $G$ is $\ell$, then the naive way of using SWAP gates (equivalent to 3 CNOT gates) would require about $6(\ell-1)$ CNOT gates (Figure \ref{fig:swapTemplate}).

Thus we devise algorithms using Steiner trees that reduce the number of CNOT gates. Steiner trees were also used in \cite{2020_NGM} and \cite{2019_KdG} for the similar goal of reducing CNOT gates.  
Our algorithms differ from these works, which we have pointed out in the following paragraphs, and give much less CNOT-count.

\begin{figure}[t!]
\renewcommand{\thefigure}{(a)}
\begin{minipage}[b]{0.44\linewidth}
\centering
\begin{tikzpicture}[scale=0.5]
\begin{scope}[every node/.style={circle,minimum size= .07 cm,draw}]
    \node (1) at (-2,2) {$\mathbf{1}$};
    \node (2) at (0,2) {$2$};
    \node (3) at (2,2) {$3$};
    \node (4) at (-2,0) {$4$};
    \node (5) at (0,0) {$5$};
    \node (6) at (2,0) {$6$};
    \node (7) at (-2,-2) {$7$};
    \node (8) at (0,-2) {$8$};
    \node (9) at (2,-2) {$\mathbf{9}$};
\end{scope}
\draw[thick] (1)--(2);
\draw (2)--(3)--(6)--(9);
\draw[thick] (9)--(8);
\draw (8)--(7)--(4)--(1);
\draw[thick] (2)--(5);
\draw (5)--(6) (4)--(5);
\draw[thick] (5)--(8);
\end{tikzpicture}   
\caption{9-qubit square grid}
  \end{minipage}\hfill
  \renewcommand{\thefigure}{(b)}
  \begin{minipage}[b]{0.54\linewidth}
    \[
\Qcircuit @C=.7em @R=.7em @!R {
          \lstick{\ket{1}} & \qswap & \qw & \qw & \qw & \qw & \qw & \qswap & \qw & \rstick{\ket{1}}\\
          \lstick{\ket{2}} & \qswap \qwx & \qw &  \qswap & \qw & \qswap & \qw  & \qswap \qwx & \qw  & \rstick{\ket{2}}\\
          \lstick{\ket{5}} & \qw & \qw & \qswap \qwx & \ctrl{1} & \qswap \qwx & \qw & \qw & \qw & \rstick{\ket{5}}\\
          \lstick{\ket{8}} & \qswap & \qw & \qw & \targ & \qw & \qw & \qswap & \qw & \rstick{\ket{8}}\\
          \lstick{\ket{9}} & \qswap \qwx & \qw & \qw & \qw & \qw & \qw & \qswap \qwx & \qw &  \rstick{\ket{1 \oplus 9}}\\
  }
\]
\caption{$CNOT_{\mathbf{1},\mathbf{9}}$ with SWAPs}
  \end{minipage}
  \renewcommand{\thefigure}{1}\caption{In the SWAP template SWAP gates are placed along the shortest path between two qubits on the given connectivity graph, in this case a 9-qubit square grid (Fig a). When the required logical qubits are on adjacent physical qubits (Fig b) then CNOT is applied. SWAP gates are again placed to get the correct logical value on all physical qubits. }
\label{fig:swapTemplate} 
\end{figure}

\subsection*{Our techniques}

Our approach can be described as \emph{slice-and-build}. Given a circuit $\ckt_I$ as a sequence of gates we slice it at ``suitable'' points and re-synthesize or build the intermediate sliced portions in a manner such that connectivity constraints are respected and at the same time we tried to reduce the number of CNOT gates. We have described two methods for \emph{slice-and-build}. 

The first procedure, CNOT-OPT-A (Algorithm \ref{alg:slice}) described in Section \ref{subsec:sliceNbuild}, has a simple slicing technique. We partition the circuit $\ckt_I$ at the position of the Hadamard ($\had$) gates. Each intermediate sub-circuit composed of the gates $\mathcal{G}_{ph}=\mathcal{G}_{univ}\setminus\{\had\}$ is re-synthesized using algorithms PHASE-NW-SYNTH (Algorithm \ref{alg:phaseNwSynth}) and LINEAR-TF-SYNTH (Algorithm \ref{alg:linTfSynth}). For each sub-circuit we first calculate the phase polynomial $\phpoly$ and overall linear transformation $\vect{A}_{slice}$. We synthesize a phase polynomial network circuit $\ckt_{ph}$ with the gates in $\mathcal{G}_{ph}$, using PHASE-NW-SYNTH (Algorithm \ref{alg:phaseNwSynth}) described in Section \ref{subsec:phaseNwSynth}. The algorithm draws inspiration from the parity network synthesis algorithm in \cite{2018_AAM}. We calculate the \emph{parity network matrix} in which each column stores a parity term. The aim is to apply a series of transformations (CNOT gates) such that each parity term occurs at least once in the circuit. Then depending on the coefficients of the parity terms we place the gates in $\mathcal{G}_{ph}\setminus\{\CNOT,\X\}$. To impose connectivity constraints we construct Steiner trees (Section \ref{subsec:steiner}) with terminals being the set of qubits (or vertices) satisfying certain conditions. Then depending on the edge information, we perform a series of CNOT operations to get the desired result. We emphasize that our way of placing CNOT gates according to the Steiner tree edges is different from that described in \cite{2020_NGM}. 
In \cite{2019_KdG} the authors remarked that Steiner trees can be used for synthesizing a circuit from its phase polynomial, but no detail was given.

The phase polynomial network corresponding to $\phpoly$ has some overall transformation $\vect{A}_{ph}$. We synthesize a circuit $\ckt_{lin}$ that implements the ``residual'' linear transformation $\vect{A}=\vect{A}_{ph}^{-1}\vect{A}_{slice}$ using LINEAR-TF-SYNTH (Algorithm \ref{alg:linTfSynth}), described in Section \ref{subsec:linTfSynth}. The main motivation of this algorithm comes from the work in \cite{2008_PMH} that synthesizes a linear reversible circuit using CNOT gates. We follow the same reverse-engineering procedure where we (i) reduce $\vect{A}$ first to an upper triangular form, (ii) transpose the result and then (iii) reduce it to a lower triangular form so that we get the identity matrix $\id$. Each linear operation corresponds to a CNOT gate and connectivity constraints are imposed by constructing series of Steiner trees. Our procedures at steps (i) and (iii) differ from the approach in \cite{2020_NGM, 2019_KdG}, and we also get less CNOT-count.

In our second procedure CNOT-OPT-B (Algorithm \ref{alg:cnotOpt}) described in Section \ref{subsec:sliceNbuild2}, the slicing points remain the $\had$ gates but the set that is partitioned is the phase polynomial $\phpoly_I$ of the entire circuit $\ckt_I$. Between two $\had$ gates (including the ends) we synthesize a phase polynomial network circuit using gates in $\mathcal{G}_{ph}$ that realizes the partial phase polynomial $\phpoly_{sub}$, comprising of terms in $\phpoly$ that become uncomputable after the $\had$ gate being placed at the end of the current slice. Here it must be noted that by the \emph{sum-over-paths} formulation (Section \ref{subsec:cktPoly}) new \emph{path variables} are introduced after application of each $\had$ gate. This renders some terms of the phase polynomial uncomputable after certain points in the circuit. The synthesis of phase polynomial network is done using PHASE-NW-SYNTH (Algorithm \ref{alg:phaseNwSynth}). Let $\vect{A}_{slice}$ be the transformation that maps the state of the qubits in $\ckt_I$ after the $\had$ gate at the beginning of a slice to the state of the qubits in $\ckt_I$ before the $\had$ gate at the end of the slice. We synthesize a circuit implementing $\vect{A}=\vect{A}_{ph}^{-1}\vect{A}_{slice}$ using LINEAR-TF-SYNTH (Algorithm \ref{alg:linTfSynth}) such that between any two $\had$ gates (as well as at the ends) the linear transformation $\vect{A}_{slice}$ remain unchanged. A similar kind of partitioning of the circuit according to the phase polynomial was used in \cite{2014_AMM} where the goal was to reduce  T-depth of the input circuit.

\subsection*{Our results}
\setcounter{figure}{1}

\begin{figure}[t!]
\centering
\begin{subfigure}{0.45\textwidth}
\centering
 \includegraphics[width=\textwidth]{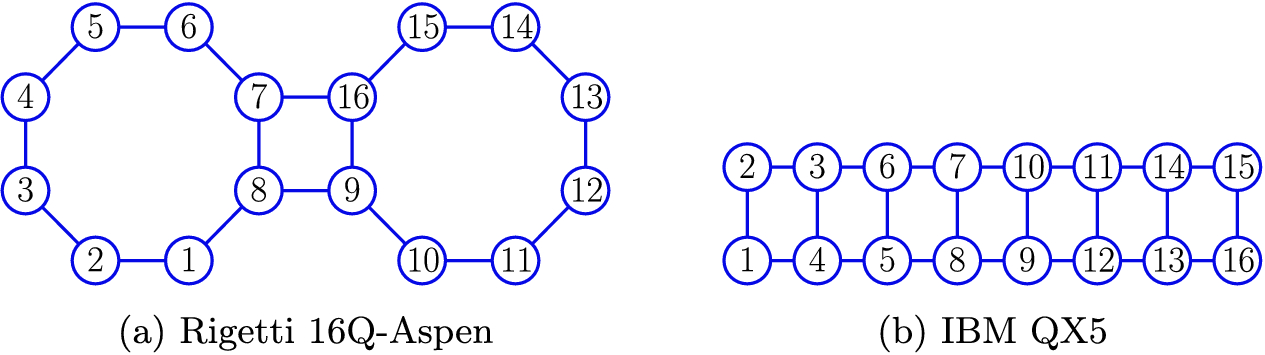}   
  \end{subfigure}
  \hfill
  \begin{subfigure}{0.45\textwidth}
  \includegraphics[width=\textwidth]{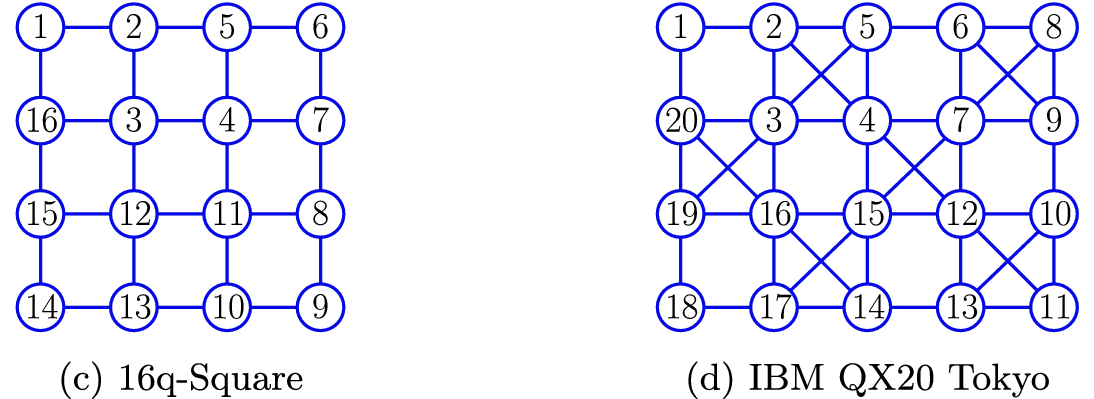}
  \end{subfigure}
   \caption{Connectivity graphs of some architectures that have been implemented in practice. Images taken from \cite{2020_dBBVMA}.}
 \label{fig:arch}
\end{figure}

We have re-synthesized some benchmark as well as randomly generated circuits after taking into account connectivity constraints imposed by the architectures (which have been implemented in practice) shown in Figure \ref{fig:swapTemplate} and \ref{fig:arch}. Here we emphasize that we have studied the performance of our procedures as baseline algorithms. The results will likely improve if coupled with some other procedures that handle the problem of optimal initial mapping of qubits. To be precise, we have considered only one mapping where qubit $i$ is mapped to vertex $i$ of the given connectivity graphs. Considerable amount of work has been done, which considers the optimal mapping that reduces the resources required. So if such a procedure is done as pre-processing, then the CNOT-count will likely reduce further. We have compared the CNOT-count overhead, that is, the increase in CNOT-count obtained from our algorithms with the overhead obtained using SWAP-template (Figure \ref{fig:swapTemplate}) and Qiskit transpiler. We have observed that both our algorithms give remarkable improvement in the case of benchmark circuits (Table \ref{tab:benchmark} in Section \ref{subsec:result}). In case of random circuits we have found that the simple way of slicing in CNOT-OPT-A gives much less overhead (Table \ref{tab:random} in Section \ref{subsec:result}). CNOT-OPT-B, however, fares poorly in many cases. 

\subsection{Related work}
 
There have been quite a number of works that deal with the problem of CNOT optimization without taking into account the connectivity constraints imposed by the underlying hardware architecture, for example, \cite{2002_IKY, 2002_SPMH, 2008_PMH, 2009_SM, 2014_oD, 2014_WGMA, 2018_AAM}. 

Some authors~\cite{2019_LDX, 2016_PS, 2016_WKWRCD, 2019_IRIMC, 2020_IRIM} use SWAP gates along with some gate commutation and transformation rules to obtain a circuit that respect connectivity constraints. There are algorithms that take advantage of the restricted topology such as 1D linear nearest neighbor (\cite{2011_MY, 2011_HNYN, 2011_SWD, 2011_CSC, 2013_SSP, 2015_RD}), hypercubic \cite{2017_B} which rely on classical sorting networks and 2D grid (\cite{2014_SSP, 2015_LWD, 2016_WKWRCD, 2017_RB}). Some algorithms that work on general topology for NISQ devices are \cite{2018_ZPW, 2017_BC, 2018_SSCP, 2019_LDX, 2019_CDDetal, 2019_WBZ}. Broadly, these algorithms use qubit mapping technique to search for the optimal placement of SWAP gates and qubits. The search space scales exponentially for exact algorithms such as~\cite{2018_VDRF, 2019_MJCM}, making them impractical for large NISQ devices. Thus, some authors~\cite{2019_CDDetal, 2019_LDX, 2021_PZW} use heuristics to reduce the search space. Some of these heuristics algorithms, e.g.~ \cite{2018_ZPW}, which is based on depth partitioning and $A^*$ search, are developed for specialized architectures such IBM devices. In \cite{2018_FA} the authors give an approach for realizing arbitrary parity-function oracles, while taking care of the underlying topology. It has been shown in \cite{2019_P} that the size of the resulting circuit is very sensitive to the original placement of the logical qubits on the device. 

To reduce CNOT-count, Steiner trees have been used in \cite{2019_KdG, 2020_NGM, 2019_WHYetal, 2020_T}, while in \cite{2020_dBBVMA} the problem is reduced to a well-known cryptographic problem - the syndrome decoding problem.  

\subsection{Organization}

After giving some preliminaries in Section \ref{sec:prelim} we describe our algorithms in Section \ref{sec:algo} and \ref{sec:algoH}. The algorithms LINEAR-TF-SYNTH and PHASE-NW-SYNTH that synthesize linear reversible circuits and phase polynomial network circuits are given in Section \ref{subsec:linTfSynth} and \ref{subsec:phaseNwSynth} respectively. The algorithms CNOT-OPT-A and CNOT-OPT-B that synthesize the complete circuit over the Clifford+T gate set is described in Section \ref{subsec:sliceNbuild} and \ref{subsec:sliceNbuild2} respectively. Finally we conclude in Section \ref{sec:conclude}.

\section{Preliminaries}
\label{sec:prelim}

We write $N=2^n$ and $[K]=\{1,2,\ldots,K\}$. The $(i,j)^{th}$ entry of any matrix $M$ is denoted by $M_{i,j}$ or $M_{ij}$ or $M[i,j]$. We denote the $i^{th}$ row of $M$ by $M[i,.]$ and the $j^{th}$ column by $M[.,j]$. We denote the $n\times n$ identity matrix by $\id_n$ or $\id$ if the dimension is clear from the context. The set of $n$-qubit unitaries of size $2^n\times 2^n$ is denoted by $\unit(2^n)$ or $\unit_n$.
\subsection{Cliffords and Paulis}

The \emph{single qubit Pauli matrices} are as follows:
\begin{eqnarray}
 \X=\begin{bmatrix}
     0 & 1 \\
    1 & 0
    \end{bmatrix} \qquad  
 \Y=\begin{bmatrix}
     0 & -i \\
     i & 0
    \end{bmatrix} \qquad 
 \Z=\begin{bmatrix}
     1 & 0 \\
     0 & -1
    \end{bmatrix}.
\label{eqn:Pauli1}
\end{eqnarray}

The \emph{$n$-qubit Pauli operators} are :
\begin{eqnarray}
 \pauli_n=\{Q_1\otimes Q_2\otimes\ldots\otimes Q_n:Q_i\in\{\id,\X,\Y,\Z\} \}.
 \label{eqn:paulin}
\end{eqnarray}

The \emph{single-qubit Clifford group} $\cliff_1$ is generated by the Hadamard and phase gate
\begin{eqnarray}
 \cliff_1=\braket{\had,\phase} 
 \label{eqn:cliff1}
\end{eqnarray}
where
\begin{eqnarray}
 \had=\frac{1}{\sqrt{2}}\begin{bmatrix}
       1 & 1 \\
       1 & -1
      \end{bmatrix}\qquad 
 \phase=\begin{bmatrix}
       1 & 0 \\
       0 & i
      \end{bmatrix}
\end{eqnarray}
When $n>1$ the \emph{$n$-qubit Clifford group} $\cliff_n$ is generated by these two gates (acting on any of the $n$ qubits) along with the two-qubit $\CNOT=\ket{0}\bra{0}\otimes\id+\ket{1}\bra{1}\otimes\X$ gate (acting on any pair of qubits). We write $\CNOT_{c,t}$ to denote the CNOT gate applied between qubit $c$ (\emph{control}) and $t$ (\emph{target}). The logic realized by this gate is : $\CNOT\ket{c,t}=\ket{c,c\oplus t}$.

The \emph{Clifford+T} gate set consists of the $n$-qubit Clifford group gates along with the $\T$ gate, where
\begin{eqnarray}
 \T=\begin{bmatrix}
     1 & 0 \\
     0 & e^{i\frac{\pi}{4}}
    \end{bmatrix}.
\end{eqnarray}
It is easy to verify that this set is a group since the $\had$ and $\CNOT$ gates are their own inverses and $\T^{-1}=\T^7$. Here we note $\phase=\T^2$. For $n>1$ qubits a minimal generating set for this group is $\{\had,\T,\CNOT\}$.

\subsection{Circuit-polynomial correspondence}
\label{subsec:cktPoly}

The circuit-polynomial correspondence \cite{2017_M} associates a \emph{phase polynomial} and a linear Boolean transformation with every quantum circuit generated by the set $\{\CNOT,\had,\T\}$. More precisely,
\begin{lemma}[\cite{2014_AMM}]
 A unitary $U\in\unit(2^n)$ is exactly implementable by an $n$-qubit circuit over $\{\CNOT,\T\}$ if and only if
 $$
    U\ket{x_1x_2\ldots x_n} = \omega^{p(x_1,x_2,\ldots,x_n)}\ket{g(x_1,x_2,\ldots,x_n)}
 $$
 where $\omega=e^{\frac{i\pi}{4}}$, $x_1,x_2,\ldots,x_n\in\fld_2$ and
 $$
    p(x_1,x_2,\ldots,x_n)=\sum_{i=1}^{\ell} c_i\cdot f_i(x_1,x_2,\ldots,x_n)
 $$
 for some linear reversible function $g:\fld_2^n\rightarrow\fld_2^n$ and linear Boolean functions $f_1,f_2,\ldots,f_{\ell}\in\left(\fld_2^n\right)^*$ with coefficients $c_1,c_2,\ldots,c_{\ell}\in\intg_8$.
 \label{lem:sop1}
\end{lemma}
For convenience, the set of $n$-ary linear Boolean functions $\fld_2^n\rightarrow\fld_2$ is referred to as the \emph{dual vector space} $\left(\fld_2^n\right)^*$ of $\fld_2^n$.

This is called the \emph{sum-over-paths} form of a circuit \cite{2005_DHMHNO, 2017_KPS, 2017_M} and the variables $x_1,x_2,\ldots,x_n$ are called the \emph{path variables}. $p(x_1,x_2,\ldots,x_n)$ is referred to as the \emph{phase polynomial}. Each $f_i(x_1,\ldots,x_n)$ is a \emph{parity term}.

Thus we can fully characterize a unitary $U\in\mathcal{U}(2^n)$ implemented by a $\{\CNOT,\T\}$-generated circuit with a set $\mathcal{P}\subseteq\intg_8\times\left(\fld_2^n\right)^*$ of linear Boolean functions together with coefficients in $\intg_8$ and a linear reversible output functions $g:\fld_2^n\rightarrow\fld_2^n$, with the interpretation
\begin{eqnarray}
 U_{\braket{\mathcal{P},g}} : \ket{x_1x_2\ldots x_n} \rightarrow \omega^{\sum_{(c,f)\in\mathcal{P}}c\cdot f(x_1,x_2,\ldots x_n)} \ket{g(x_1,x_2,\ldots,x_n)}.
 \label{eqn:phasePoly}
\end{eqnarray}
The set $\mathcal{P}$ (\emph{phase polynomial set}) and $g$ are efficiently computable given a circuit over $\{\CNOT,\T\}$, taking time linear in the number of qubits and gates. 

The $\had$ gate is a ``branching gate'' and has the following effect on a basis state $x_1\in\fld_2$.
\begin{eqnarray}
 H:\ket{x_1}\rightarrow\frac{1}{\sqrt{2}} \sum_{x_2\in\fld_2}\omega^{4\cdot x_1\cdot x_2}\ket{x_2}  \nonumber
\end{eqnarray}
Here $x_2$ is the new path variable and the variable $x_1$ ceases to exist after $\had$ is applied. Similar to Lemma \ref{lem:sop1} we can have the following result.
\begin{lemma}[\cite{2014_AMM}]
If a unitary $U\in\unit(2^n)$ is exactly implementable by an $n$-qubit circuit over $\{\CNOT,\had,\T\}$ with $k$ $\had$ gates, then for $x_1x_2\ldots x_n\in\fld_2^n$,
$$
    U\ket{x_1x_2\ldots x_n}=\frac{1}{\sqrt{2^k}}\sum_{x_{n+1}\ldots x_{n+k}\in\fld_2^k} \omega^{p(x_1,x_2,\ldots,x_{n+k})} \ket{y_1y_2\ldots y_n}
$$
where $y_i=h_i(x_1,x_2,\ldots,x_{n+k})$ and
$$
    p(x_1,x_2,\ldots,x_{n+k})=\sum_{i=1}^{\ell}c_i\cdot f_i(x_1,\ldots,x_{n+k})+4\cdot \sum_{i=1}^{k} x_{n+i}\cdot g_i(x_1,\ldots,x_{n+k})
$$
for some linear Boolean functions $h_i,f_i,g_i$ and coefficients $c_i\in\intg_8$. The $k$ path variables $x_{n+1},\ldots,x_{n+k}$ result from the application of Hadamard gates.
 \label{lem:sop2}
\end{lemma}
But, unlike Lemma \ref{lem:sop1}, the converse is not true.

\subsection{Steiner tree}
\label{subsec:steiner}

A \textbf{graph} is a pair $G=(V_G,E_G)$ where $V_G$ is a set of \emph{vertices} and $E_G$ is a set of pairs $e=(u,v)$ such that $u,v\in V_G$. Each such pair is called an \emph{edge}. One may define a function $w_{E_G}:E_G\rightarrow\real$ that assigns a \emph{weight} to each edge. 
The graphs we consider are \emph{simple} (with at most one edge between two distinct vertices and no self-loops i.e. $(u,u)\notin E_G$), \emph{undirected} (edges have no direction i.e. $(u,v)\equiv (v,u)$) with edge-weight 1, that is, $w_{E_G}(e)=1$ for every $e\in E_G$. A graph $G'=(V_G^{'},E_G^{'})$ is a \emph{subgraph} of $G$ such that $V_G^{'}\subseteq V_G$ and $E_G^{'}\subseteq E_G$. A \textbf{tree} is an undirected graph in which any two vertices are connected by exactly one path, or equivalently a connected, acyclic, undirected graph. 

\begin{definition}[\textbf{Steiner tree}]
Given a graph $G=(V_G,E_G)$ with a weight function $w_E$ and a set of vertices $\tset\subseteq V_G$, a Steiner tree $T=(V_T,E_T)$ is a minimum weight tree that is a subgraph of $G$ such that $\tset\subseteq V_T$. 

The set of vertices in $\tset$ are called \emph{terminals} while those in $V_T\setminus \tset$ are called \emph{Steiner nodes}.
 \label{defn:steinerTree}
\end{definition}

Computing Steiner trees is NP-hard and the related decision problem is NP-complete \cite{1972_K}. There are a number of heuristic algorithms that compute approximate Stiener trees~\cite{1992_HR, 2005_RZ, 2013_BGRS}. The choice of algorithm is usually determined by the application, and typically involves a trade-off between quality (approximation factor) and running time. We use the algorithm given in \cite{2013_SF} (Algorithm \ref{alg:steinerTree}), incorporating optimization steps by Rayward et al.\cite{1986_RC}. This helps us achieve a better running time compared to the best (with respect to quality) approximation algorithms in the literature, without sacrificing the approximation factor much. The primary idea of the algorithm is to maintain a number of subgraphs and sequentially merge those which are closest to each other. The distance between two subgraphs $g_i,g_j$ is measured by the length of the shortest path between any two nodes $u,v$ such that $u\in V_{g_i}\setminus V_{g_j}$ and $v\in V_{g_j}\setminus V_{g_i}$. When a subgraph has all terminals then we stop the merging and remove all non-terminal nodes of degree 1. A pseudocode of this algorithm has been given in Appendix \ref{app:steinerTree} (Algorithm \ref{alg:steinerTree}).

The size of the constructed Steiner tree is at most $2\left(1-\frac{1}{\ell}\right)$ times the size of the minimal Steiner tree, where $\ell$ is the number of leaves in the minimal Steiner tree. The running time is $O\left(|\tset|^2\left(|V_G|+|E_G| \right) \right)$ \cite{2013_SF}.


\section{Synthesis algorithms with connectivity constraint}
\label{sec:algo}

In this section first we describe a synthesis algorithm that generates a circuit implementing a linear transformation using gates in the set $\mathcal{G}_{lin}=\{\CNOT,\X\}$ (Section \ref{subsec:linTfSynth}). Then we describe an algorithm that synthesizes a circuit implementing a phase polynomial network using gates $\mathcal{G}_{ph}$ generated by the set $\{\CNOT,\X,\T\}$.

\subsection{Synthesis of $\{\CNOT,\X\}$ circuits }
\label{subsec:linTfSynth}

\begin{algorithm}
\footnotesize
\caption{LINEAR-TF-SYNTH}
\label{alg:linTfSynth}
 
 \KwIn{(i) Linear transformation matrix $\vect{A}_{n\times n+1}=[\vect{A}^{'}_{n\times n}|\vect{b}_{n\times 1}]$, (ii) Connectivity graph $G$.}
 
 \KwOut{A circuit over $\{\CNOT,\X\}$ realizing $\vect{A}$.}
 
 $\mathcal{Y}_1,\mathcal{Y}_2,\mathcal{X}\leftarrow\emptyset$, $G'\leftarrow G$ \tcp*[f]{Make a copy of $G$}   \;
 
 If $A_{n+1,i}=1$ then $\mathcal{X}.append(\X_i)$ and $A_{n+1,i}\leftarrow 0$ \label{linTf:X}\;
 
 \For{ columns $i=1\ldots,n$}
 {
    If $A_{ii}=0$, find all rows $j (j>i)$ such that $A_{ji}=1$. Choose $j$ with the shortest path (in $G'$) to $i$. Suppose $i$ is the root and $j$ is the leaf. $\mathcal{Y}_1.append(\CNOT_{uv})$ if $u$ is the child of $v$. $\vect{A}[v,.]\leftarrow\vect{A}[v,.]\oplus\vect{A}[u,.] $ \; \label{linTf:Aii}
    
    $\tset'=\{j: j>i \text{ and } A_{j,i}=1 \}$, $\tset\leftarrow \tset'\cup\{i\}$. 
     \tcp*[f]{$\tset$ is the set of terminals.} \;
     
     Find a minimum Steiner tree approximation with connectivity graph $G'$ and terminals $\tset$. Let $i$ be the root of this tree, $T_{i,\tset}$. \;
     
     $\mathcal{Y}\leftarrow\emptyset$; $\quad alg\leftarrow 1$  \; \label{linTf:utY}
     
     $(\mathcal{Y},\vect{A})\leftarrow$ ROW-OP($\vect{A}, \tset, i, T_{i,\tset},alg$). \label{linTf:uTri} \;
     
     $\mathcal{Y}_1.append(\mathcal{Y})$, $G'\leftarrow G'\setminus\{i\}$ \tcp*[f]{Remove vertex $i$ and its edges from $G'$.} \;
 }
 
 Transpose $\vect{A}$ and $G'\leftarrow G$. \label{linTf:transpose} \;
 \For{ columns $i=1\ldots,n$}
 {
    Repeat steps 3-6 \;
    
    $\mathcal{Y}\leftarrow\emptyset$; $\quad alg\leftarrow 2$  \;
    
    $(\mathcal{Y},\vect{A},\mathcal{T}_1)\leftarrow$ ROW-OP($\vect{A}, \tset, i, T_{i,\tset},alg$) \label{linTf:lTri} \;
    
    $\mathcal{Y}_2.append(\mathcal{Y})$ \;
    
    Initialize an array $B$ of size $n$ and $B[\ell]\leftarrow r$ if $\ell\in \tset\setminus\{i\}$ and $r$ is the root of the sub-tree in which $\ell$ was a leaf   \label{linTf:B}\;
    
    $t\leftarrow |\mathcal{T}_1|$   \;
    
    \For{$j=0,\ldots t-1$}
    {
        $(r,\ell_1,\ldots,\ell_m)\leftarrow\mathcal{T}_1[j]$    \;
        
        \While{$\exists k \text{ such that } \ell_k < r$  \label{linTf:corrWhile}}
        {
            $\ell\leftarrow \ell_k$ \;
            
            \While{$r > \ell$}
            {
                $\tset\leftarrow\{r,\ell\}$; $\quad T_{r,\tset} \leftarrow $ The shortest path between $r$ and $\ell$ stored as tree with root $r$ and leaf $\ell$; $\quad\mathcal{Y}\leftarrow\emptyset$; $\quad alg\leftarrow 3$  \;
                
                $(\mathcal{Y},\vect{A})\leftarrow$ ROW-OP($\vect{A},\tset,r,T_{r,\tset},alg$) \;
            
                $\mathcal{Y}_2.append(\mathcal{Y})$ \;
                
                $B[\ell]=B[r]$, $r\leftarrow B[r]$ \;
            }
        }
        \label{linTF:corrWhileEnd}
        
    }
    $G'\leftarrow G'\setminus\{i\}$ \;
 }
 
 $\mathcal{Y}_2^{'}=\mathcal{Y}_2$ with control and target flipped for each CNOT gate \;
 
 \Return $\mathcal{Y}_2^{'}\cup reverse(\mathcal{Y}_1)\cup\mathcal{X}$ as the circuit. 
 
\end{algorithm}

Consider an $n$-qubit circuit built with gates in the set $\mathcal{G}_{lin}=\{\CNOT,\X\}$. We represent the overall linear transformation by an $n\times n+1$ ``augmented'' matrix $\vect{A}=[\vect{A}_{n\times n}^{'}|\vect{b}_{n\times 1}]$, whose rows represent or are indexed by qubits. If we label the initial states of the qubits by variables $x_1,\ldots,x_n$ then the first $n$ columns represent these variables and the last column represent the variable $b$ indicating bit flips. Each variable $x_1,\ldots,x_n,b$ takes values from the set $\{0,1\}$. The initial state of $\vect{A}$ is $[\id_n|\vect{0}_{n\times 1}]_{n\times n+1}$. This represents the initial state of all the qubits. When $\CNOT_{j,i}$ is applied row $j$ is added (mod 2) to row $i$ (row $j$ remains same). The parity at qubit $i$ is $x_i\oplus x_j$. When an $\X$ gate is applied on qubit $i$ then  $A_{i,n+1}\leftarrow 1\oplus A_{i,n+1}$.

Now suppose we are given a linear transformation $\vect{A}=[\vect{A}_{n\times n}^{'}|\vect{b}_{n\times 1}]$ of a circuit and we want to synthesise a circuit implementing this transformation. We use the same reverse engineering idea of Patel, Markov and Hayes \cite{2008_PMH}. The procedure is similar to Gaussian elimination. (a) First we make $\vect{b}=\vect{0}$ by flipping the entries with 1. This corresponds to applying X on the respective qubit. (b) We apply a series of elementary row operations (bit-wise addition) on $\vect{A}$ such that $\vect{A}'$ is in upper triangular form. Each row operation represents the application of a $\CNOT$ gate. (c) Then we transpose the matrix and perform elementary row operations on $\vect{A}^T$ such that $\vect{A}'$ is $\id$. The output circuit is constructed as follows : first, the $\CNOT$ gates obtained in (c) with the control-target flipped but preserving the order, then the CNOT gates obtained in (b) with control-target preserved but reversing the order in which they were performed, and lastly the X gates obtained in (a).

To incorporate connectivity constraints we use Steiner trees as described in LINEAR-TF-SYNTH (Algorithm \ref{alg:linTfSynth}).
We first make $\vect{b}=\vect{0}$ by placing X gates (step \ref{linTf:X}), as described before. Then we convert $\vect{A}'$ into an upper triangular form (step \ref{linTf:uTri}) by row-operations ``permitted'' by the input connectivity graph $G$. This is a graph whose vertices represent qubits and a two-qubit gate such as CNOT can be placed only when there exists an edge between the corresponding vertices. For each column of $\vect{A}'$ (starting from the first one) we compute a minimal Steiner tree approximation with (i) connectivity graph $G'=G\setminus I$ (excluding the vertices in $I$ and the edges adjacent to these vertices) where $I$ is the set of columns which have been operated on or which have been ``fixed'' to have 1 in the diagonal and 0 in the rest, and (ii) set of terminals $\tset$ which are the rows below the diagonal and having a $1$. Then we invoke the procedure ROW-OP, as described in Algorithm \ref{alg:rowOP}.

\begin{algorithm}
\footnotesize
 \caption{ROW-OP}
 \label{alg:rowOP}
 
 \KwIn{(i) Linear transformation matrix $\vect{A}$, (ii) Set of terminals $\tset$, (iii) Pivot node $c$, (iv) A minimal Steiner tree approximation $T_{c,\tset}$, (v) $alg \in \intg$. }
 
 \KwOut{(i) List $\mathcal{Y}$ of $\CNOT$ operations, (ii) $\vect{A}$ (after the row operations), (iii) $\mathcal{T}_1=\{(r,\ell_1,\ldots,\ell_m): r \text{ is the root and } \ell_1,\ldots,\ell_m \text{ are the leaves in a sub-tree}\}$ if $alg==2$. }
 
  $\mathcal{T}\leftarrow$SEPARATE($T_{c,\tset}, c, \tset,alg$) \tcp*[f]{Separate $T_{c,\tset}$ into a set $\mathcal{T}$ of sub-trees $T_{k,\tset_k}$ with root $k\in \tset$ and leaves $\tset_k\setminus\{k\}\subset \tset$.} \label{rowOP:separate}  \;
  
  $\mathcal{T}_1=\{(r,\ell_1,\ldots,\ell_m): r \text{ is the root and } \ell_1,\ldots,\ell_m \text{ are the leaves in a sub-tree}\}$ \;
  
  $t\leftarrow\left|\mathcal{T} \right|$; $\quad\mathcal{Y}\leftarrow\emptyset$   \;
  
  \For{$i=t-1,t-2,\ldots,0$ \tcp*[f]{Starting from the last sub-tree}   \label{rowOp:4stepStart}}
  {
    
    \If{$alg\neq 1$}
    {
        (\emph{Bottom-Up-1}:) Starting from the last layer till layer $1$, $\mathcal{Y}.append(\CNOT_{uv})$ if $u$ is a non-root parent node and $v$ is a child of $u$. \label{rowOP:botUp1}  \;
        \If{$alg\neq 4$}
        {
            $\vect{A}[v,.]\leftarrow\vect{A}[v,.]\oplus\vect{A}[u,.]. $ \label{rowOP:botUp1A}    \; 
        }
    }
    
    (\emph{Top-Down1}:) Starting from top layer till last layer, $\mathcal{Y}.append(\CNOT_{uv})$ if $u$ is parent of $v$.  \label{rowOP:topD}   \;
    \If{$alg\neq 4$}
        {
            $\vect{A}[v,.]\leftarrow\vect{A}[v,.]\oplus\vect{A}[u,.]. $ \label{rowOP:topD1A}    \; 
        }
        
    (\emph{Bottom-Up-2}:) Starting from the last layer till layer $0$, $\mathcal{Y}.append(\CNOT_{uv})$ if $u$ is a parent node and $v$ is a non-leaf child node of $u$.  \label{rowOP:botUp2} \;
    \If{$alg\neq 4$}
        {
            $\vect{A}[v,.]\leftarrow\vect{A}[v,.]\oplus\vect{A}[u,.]. $ \label{rowOP:botUp2A}    \; 
        }
    
    \If{$alg\neq 1$}
    {
        (\emph{Top-Down-2}:) Starting from the second layer till last layer, $\mathcal{Y}.append(\CNOT_{uv})$ if $u$ is a parent node (that is not a child of the root) and $v$ is a non-leaf child of $u$.  \label{rowOP:topD2}    \; 
        \If{$alg\neq 4$}
        {
            $\vect{A}[v,.]\leftarrow\vect{A}[v,.]\oplus\vect{A}[u,.]. $ \label{rowOP:topD2A}    \; 
        }
    }
    \If{$alg==4$}
    {
        $\vect{A}[r,.]\leftarrow\vect{A}[r,.]\oplus\vect{A}[\ell,.]$, where $r,\ell$ are the root and leaf of the current sub-tree respectively.  \label{rowOp:A4}   \;
    }
   \label{rowOp:4stepEnd} 
  }

 \eIf{$alg\neq 2$}
 {
    \Return $(\mathcal{Y},\vect{A})$    \;
 }
 {
    \Return $(\mathcal{Y},\vect{A},\mathcal{T}_1)$ \;
 } 

\end{algorithm}

The idea of ROW-OP is to use a set of operations such that 1 in the diagonal is ``propagated'' via intermediate Steiner nodes to cancel the 1 in the terminal nodes and then use another set of operations to cancel any 1s in the Steiner nodes. We assume the diagonal has 1, else it is adjusted by a set of operations to propagate a 1 to the diagonal node (step \ref{linTf:Aii} of Algorithm \ref{alg:linTfSynth}). The diagonal node (let's call it $c$) becomes the ``pivot'' node. The input Steiner tree approximation $T_{c,\tset}$ is separated into a set of sub-trees (step \ref{rowOP:separate} of Algorithm \ref{alg:rowOP}) by calling the procedure SEPARATE (Algorithm \ref{alg:separate}). The root and leaves in each such sub-tree are terminal nodes (from $\tset$) and the rest are Steiner nodes. Then the 1 from the root of each sub-tree cancels the 1 at the leaves via operations performed in steps \ref{rowOP:topD},\ref{rowOP:topD1A} of Algorithm \ref{alg:rowOP} and the 1s at Steiner nodes get cancelled by the operations performed in steps \ref{rowOP:botUp2}, \ref{rowOP:botUp2A} of the same procedure. If in a sub-tree the root node is $r$ and the leaves are $\ell_1,\ldots,\ell_m$ then the parity at the root node and each Steiner node remains unchanged but the parity at leaf $\ell_i$ become $x_{\ell_i}\oplus x_r \bigoplus_{j\in P} x_j$ where $P$ is the set of Steiner nodes in the path from $r$ to $\ell_i$. The resultant matrix $\vect{A}'$ is in upper triangular form.

Next we transpose $\vect{A}'$. Our goal is now to convert $\vect{A}'^T$ into upper triangular form without destroying the $0$s in the upper-triangle. This in turn implies that for each non-diagonal node $j$ we want the parity to be $x_j'\oplus x_k'$, where $k<j$ and $x_j', x_k'$ are the parities at node $j$ and $k$ respectively before the transpose step \ref{linTf:transpose} in Algorithm \ref{alg:linTfSynth}. Similarly as before, we invoke the procedure ROW-OP (Algorithm \ref{alg:rowOP}) but this time we include steps \ref{rowOP:botUp1},\ref{rowOP:botUp1A} and \ref{rowOP:topD2},\ref{rowOP:topD2A} in it, so that for each sub-tree constructed the parity at root $r$ and Steiner nodes remain unchanged, but the parity at each leaf node $\ell$ becomes $x_r'\oplus x_{\ell}'$. Now if $r>\ell$ then we perform some correction procedures (step \ref{linTf:corrWhile}-\ref{linTF:corrWhileEnd} in Algorithm \ref{alg:linTfSynth}). Note the parity at $r$ is $x_{r_i}'\oplus x_r'$ where $r_i$ is the root of the sub-tree in which $r$ was a leaf. Then if we invoke ROW-OP with the shortest path from $r$ to $\ell$ as a tree, then the parity at $\ell$ becomes $x_{r_i}'\oplus x_{\ell}'$. Every other parity remains unaffected. If $r_i>\ell$, then we again invoke ROW-OP with the shortest path from $r_i$ to $\ell$ as a tree. We continue doing this till the parity at $\ell$ is ``corrected'' i.e. it becomes $x_{\ell}'\oplus x_k'$ for some $k<\ell$. We start these correction procedures from the first sub-tree, so we can guarantee that the parity at each node gets corrected as desired.

\begin{algorithm}
\footnotesize
 \caption{SEPARATE}
 \label{alg:separate}
 
 \KwIn{(i) Steiner tree $T_{c,\tset}$, (ii) Pivot $c$, (iii) Set of terminals $\tset$. (iv) $alg\in\intg$.}
 
 \KwOut{$\mathcal{T}=\{T_{k,\tset_k}: \text{ Edge-disjoint sub-trees with root } k\in \tset \text{ and leaves } \tset_k\setminus\{k\}\subset \tset \}$.}
 
 $\mathcal{T}\leftarrow\emptyset$, $root\leftarrow c$, $R\leftarrow\{root\}$, $\tset.delete(c)$ \;
 
 \While{$\tset\neq\emptyset$}
 {
    $\tset_{root}\leftarrow\{root\}$, $T_{root,\tset_{root}}\leftarrow\emptyset$ \tcp*[f]{Initialize the data-structure to store a sub-tree} \;
    
    Starting from $root$, traverse $T_{c,\tset}$ in breadth first search order. Store the vertices and edges in $T_{root,\tset_{root}}$ \;
    
    When arriving at a non-leaf terminal $u$, then $\tset_{root}.append(u)$ and store $u$ as a leaf in $T_{root,\tset_{root}}$, $\tset.delete(u)$, $R.append(u)$ \;
    
    \eIf{$alg\neq 4$}
    {
        $\mathcal{T}.append(T_{root,\tset_{root}})$ \tcp*[f]{Store the tree-information depth-wise.}\;
    }
    {
        \For{$u\in \tset_{root}\setminus\{root\}$ \label{separate:pathStart}}
        {
            $\tset_{u}\leftarrow\{u,root\}$ \;
            $T_{u,\tset_u}\leftarrow$ Path from $u$ to $root$ \tcp*[f]{Store the tree as if $u$ is root and $root$ is leaf.}  ;
            $\mathcal{T}.append(T_{u,\tset_u})$    \;
            \label{separate:pathEnd}
        }
    }
    
    $R.delete(root)$, $root\leftarrow R[0]$ \;
 }
 \Return $\mathcal{T}$  \;
 
\end{algorithm}

\subsubsection*{Remark \ref{subsec:linTfSynth}}

The use of Steiner trees to take care of connectivity constraints was also done in \cite{2020_NGM} and \cite{2019_KdG}. Our procedures are different from both of them. While calling the procedure ROW-OP during the reduction to upper-triangular form (before transpose in step \ref{linTf:transpose} in Algorithm \ref{alg:linTfSynth}) we skipped some steps (steps \ref{rowOP:botUp1},\ref{rowOP:botUp1A} and \ref{rowOP:topD2},\ref{rowOP:topD2A} in Algorithm \ref{alg:rowOP}) because it was not necessary and this reduced the CNOT count. We traverse each Steiner tree twice, so the number of CNOT gates required is approximately $2e$ where $e$ is the number of edges in the tree. In contrast the algorithm in \cite{2020_NGM} in this phase consumes approximately $4e$ CNOT gates. After transposing in step \ref{linTf:transpose} in Algorithm \ref{alg:linTfSynth} our procedure is markedly different from the approach taken in \cite{2020_NGM}. Even our ``correction procedure'' is different from the recursive approach taken in \cite{2019_KdG} for general graphs. Asymptotically the complexity of LINEAR-TF-SYNTH is similar to the corresponding algorithms in \cite{2020_NGM} and \cite{2019_KdG}. There are $n$ Steiner trees constructed for each of the $n$ columns. Each Steiner tree approximation will always be of size $O(n)$. The number of CNOT gates applied is $O(n)$. So overall complexity is $O(n^2)$.
An illustration of LINEAR-TF-SYNTH has been shown in Appendix \ref{app:lin}, using an example given in \cite{2020_NGM}. We re-synthesize a given linear transformation circuit using 26 CNOTs, while \cite{2020_NGM} used 43 CNOTs for re-synthesizing the same circuit.

\subsection{Synthesis of circuits over $\{\CNOT,\X,\T\}$}
\label{subsec:phaseNwSynth}

\begin{algorithm}
\footnotesize
 \caption{PHASE-NW-SYNTH}
 \label{alg:phaseNwSynth}
 
 \KwIn{(i) $\mathcal{P}\in\{(c,f):c\in\intg_8\text{ and }f\in\left(\fld_2^n\right)^*\times\fld_2\}$, (ii) Connectivity graph $G$.}
 \KwOut{A circuit with gates in $\mathcal{G}_{ph}=\{\CNOT,\X,\T,\T^{\dagger},\phase,\phase^{\dagger},\Y,\Z\}$ realizing the phase polynomial network given by $\phpoly$.}
 
 $\mathcal{Y}'\leftarrow\emptyset$; $\quad I\leftarrow [n]$; $\quad \stack\leftarrow\emptyset$ \tcp*[f]{Initialize an empty stack $\stack$.}  \;
 If $(c_i, x_i) \in \mathcal{P}$ or $(c_i, 1\oplus x_i)\in\mathcal{P}$ for $i\in[n]$ then $\mathcal{Y}'.append(U[i])$ or $\mathcal{Y}'.append(\X[i] U[i])$ respectively, where $U[i]\in\mathcal{G}_{ph}$ is determined by $c_i$. Then delete these terms from $\mathcal{P}$.  \; 
 Construct the parity matrix $\vect{P}_{n+2\times p}$ where $p=$ number of terms in $\mathcal{P}$, the first $n$ rows of each column is a parity term (without bit flip term), $n+1^{th}$ row stores the bit flip and $n+2^{th}$ row stores the coefficients \;
 $B\leftarrow\{\vect{p}_i:\vect{p}_i=[(\vect{P}[0:n-1,i])^T||i]^T \}$ \tcp*[f]{$\vect{p}_i$ is the first $n$ rows of $i^{th}$ column of $\vect{P}$, appended by $i$, the column index.}     \label{phaseNw:B}\;
 \If{$B\neq\emptyset$}
 {
    $\stack.push(B,I,\epsilon)$ \;
 }
 \While{$\stack\neq\emptyset$}
 {
    $(B,I,i)\leftarrow \stack.pop()$ \;
        \If{$i\in\nat$}
        {
            $\tset'\leftarrow \{k\in[n]: k\neq i \text{ and } p_k=1  \quad \forall\vect{p}\in B \}$ \label{phaseNw:S'}\;
            \If{$\tset'\neq\emptyset$}
            {
                $\tset=\tset'\cup\{i\}$ ;  $\quad alg\leftarrow 4$   \;
                Find a minimum Steiner tree approximation with connectivity graph $G$ and terminals $\tset$. Let $i$ be the root of this tree, $T_{i,\tset}$  \label{phaseNw:steiner}\;
                $\vect{A}=$ Matrix with columns $\vect{p}\in B'$ such that $(B',I',i')\in \stack\cup (B,I,i)$    \label{phaseNw:A}\;
                $\quad\mathcal{Y}\leftarrow\emptyset$   \;
                $(\mathcal{Y},\vect{A})\leftarrow$ ROW-OP$(\vect{A},\tset,i,T_{i,\tset},4$)    \;
                $\mathcal{Y}'.append(\mathcal{Y})$  \;
                \If{$\exists \vect{p}\in B'$ where $(B',I',i')\in \stack\cup (B,I,i)$ such that $p_k=0$ for $k\in [n]$ and $k\neq i$ \label{phaseNw:BdelStart}}
                {
                    At the place in the circuit where the parity given by $p_n^{th}$ column of $\vect{P}$ is realized, place (append in $\mathcal{Y}'$) $\X U$ or $U$ (accordingly)    \;
                    $B'.delete(\vect{p})$    \;
                }
                \label{phaseNw:BdelEnd}
            }
        }
        $j\leftarrow arg \max_{j\in I} \max_{x\in\fld_2}\left|\{\vect{p}\in B : p_j=x \}\right|$    \label{phaseNw:pivot}\;
        $B_0\leftarrow\{\vect{p}\in B : p_j=0 \}$;
        $\quad B_1\leftarrow\{\vect{p}\in B : p_j=1 \}$   \label{phaseNw:B01}\;
        \If{$B_1\neq\emptyset$}
        {
            \eIf{$i=\epsilon$}
            {
                $\stack.push(B_1,I\setminus\{j\},j)$ \;
            }
            {
                $\stack.push(B_1,I\setminus\{j\},i)$ \;
            }
        }
        \If{$B_0\neq\emptyset$}
        {
            $\stack.push(B_0,I\setminus\{j\},i)$ \;
        }
 }
 \Return $\mathcal{Y}'$ \;
 
\end{algorithm}

We consider the circuits implemented with the set of gates ($\mathcal{G}_{ph}$) generated by $\{\CNOT,\X,\T\}$. Since $\phase=\T^2,\quad \Z=\T^4, \quad\T^{\dagger}=\T^7$ and $\phase^{\dagger}=\T^6$, so $\mathcal{G}_{ph}=\{\CNOT,\X,\T,\T^{\dagger},\phase,\phase^{\dagger},\Y,\Z\}$. We know from Lemma \ref{lem:sop1} in Section \ref{subsec:cktPoly} that a unitary implemented over $\{\CNOT,\T\}$ can be characterized by a set $\mathcal{P}=\{(c,f):c\in\intg_8\text{ and } f\in\left(\fld_2^n\right)^*\}$ and linear reversible output functions $g:\fld_2^n\rightarrow\fld_2^n$ (Equation \ref{eqn:phasePoly}). This actually holds for circuits over $\{\CNOT,\X,\T\}$.

Given an $n$-qubit circuit over $\{\CNOT,\X,\T\}$ with input path variables $x_1,x_2,\ldots,x_n$, we can compute each $\mathcal{P}$ as follows: For each gate $U\in\{\T,\T^{\dagger},\phase,\phase^{\dagger},\Z,\Y\}$ consider the parity, $\bigoplus_{j\in \tset} x_j\oplus b$ for $\tset\subseteq [n]$, of the qubit just before $U$ acts. Here $b\in\{0,1\}$ is the \emph{bit variable} that takes the value $1$ only after an X or Y gate acts. This is represented by the function $f$. The coefficient $c$ is given by $\{1,7,2,6,4,4,4\}$ respectively. For $(c_1,f_1),(c_2,f_2)\in\mathcal{P}$ if $f_1=f_2=f$ then we can merge them into a single pair $(c_1+c_2\mod 8, f)$.

The linear reversible output function is $g:\fld_2^n\times\fld_2\rightarrow\fld_2^n\times\fld_2$ (one of the variables is the bit flip variable $b$). More detail about the matrix representing $g$ and procedures to synthesize circuits over $\{\CNOT,\X\}$ that realize $g$ has been given in Section \ref{subsec:linTfSynth}.

We follow the approach taken in \cite{2018_AAM} and \cite{2020_NGM} while re-synthesizing circuits over $\{\CNOT,\X,\T\}$. Both these authors consider a restricted gate set consisting of $\CNOT$ and rotation gates $R_Z$. Given a phase polynomial set $\mathcal{P}$ and matrix $\vect{A}$ corresponding to the linear reversible output function $g$, they first synthesized a \emph{parity network} (defined below) that realizes the parity terms ($f$ where $(c,f)\in\mathcal{P}$) in $\mathcal{P}$. Then they applied the rotation gates depending on the coefficients ($c$) in $\mathcal{P}$. After that they synthesized a circuit such that the overall linear transformation is $\vect{A}$. While the algorithm in \cite{2020_NGM} takes care of connectivity constraints, the one in \cite{2018_AAM} is oblivious to it.

\begin{definition}[\textbf{Parity network}]

A parity network for a set $\mathcal{P}=\{(c,f):c\in\intg_8\text{ and } f\in\left(\fld_2^n\right)^*\times\fld_2\}$ is an $n$-qubit circuit over $\{\CNOT,\X\}$ gates in which each \emph{parity} term $f$ such that  $(c,f)\in\mathcal{P}$ appears at least once.

 \label{defn:parityNw}
\end{definition}

\begin{definition}[\textbf{Phase polynomial network}]

A phase polynomial network for a set $\mathcal{P}=\{(c,f):c\in\intg_8\text{ and } f\in\left(\fld_2^n\right)^*\times\fld_2\}$ is an $n$-qubit circuit over $\{\CNOT,\X,\T\}$ such that for each element $(c,f)\in\mathcal{P}$ the parity $f$ appears before a gate in $\{\T,\T^{\dagger},\phase,\phase^{\dagger},\Z,\Y\}$ when $c\in\{1,7,2,6,4,4\}$ respectively.

 \label{defn:phaseNw}
\end{definition}

We now describe our algorithm PHASE-NW-SYNTH (Algorithm \ref{alg:phaseNwSynth}) that synthesizes a phase polynomial network given by $\mathcal{P}$. We construct the parity network matrix $\vect{P}$, which has $n$ rows corresponding to each qubit and where each column corresponds to a parity term $f$ in $\mathcal{P}$. Similar to \cite{2018_AAM}, the optimization procedure to synthesize the parity network represented by $\vect{P}$ is inspired by \emph{Gray codes} \cite{1953_F}, which cycle through the set of $n$-bit strings using the exact minimal number of bit flips.
Given a set $B$ of binary strings (step \ref{phaseNw:B}), we synthesize a parity network for $B$ by repeatedly choosing an index $j$ (step \ref{phaseNw:pivot}) to expand and then effectively recurring on the co-factors $B_0$ and $B_1$ (step \ref{phaseNw:B01}), consisting of the strings $\vect{p}\in B$ with $p_j=0$ or $1$ respectively. As a subset $B$ is recursively expanded, CNOT gates are applied so that a designated \emph{target} qubit $i$ contains the partial parity $\bigoplus_{k\in \tset'} x_k$ where $\tset'$ is the set of qubits (or row indices) such that $p_k=1$ ($k\neq i$) for all $\vect{p}\in B$ (step \ref{phaseNw:S'}). Whenever a column has a single 1, it implies that the corresponding parity has been realized. So we can remove these columns from the set $B'$ of ``remaining parities'' (steps \ref{phaseNw:BdelStart}-\ref{phaseNw:BdelEnd}). At this step we can place the gate $\X$ if parity realized on circuit is $1\oplus f$ for some $(c,f)\in\mathcal{P}$. We can also place a gate in $\{\T,\T^{\dagger},\phase,\phase^{\dagger},\Z,\Y\}$ corresponding to the value of the coefficient $c$.

To incorporate connectivity constraints we find a minimal Steiner tree $T_{i,\tset}$ with connectivity graph $G$ and terminals $\tset=\tset'\cup\{i\}$ (step \ref{phaseNw:steiner}). We call the procedure ROW-OP (Algorithm \ref{alg:rowOP}) with the matrix $\vect{A}$ such that its columns are the set of unrealized parities \ref{phaseNw:A}. ROW-OP calls the sub-routine SEPARATE \ref{alg:separate} which as before separates $T_{i,\tset}$ into edge-disjoint sub-trees such that in each tree the root and leaves belong to set of terminals $\tset$. However, unlike the previous methods, this time each sub-tree with multiple leaves are further sub-divided such that each tree has a single leaf. Each such tree is stored in reverse depth-first order such that the leaf becomes root and vice-versa (steps \ref{separate:pathStart}-\ref{separate:pathEnd} in Algorithm \ref{alg:separate}). Now when we perform steps \ref{rowOp:4stepStart}-\ref{rowOp:4stepEnd} of Algorithm \ref{alg:rowOP} then for each sub-tree the parity at root is $x_r\oplus x_{\ell}$ where $r,\ell$ are the root and leaf of the sub-tree respectively (before flipping). Now since we process the trees from last sub-tree to first, so the net parity at node (or qubit $i$) is $\bigoplus_{k\in \tset'} x_k$. To maintain the invariant that the remaining parities are expressed over the current state of the qubits, we modify the matrix $\vect{A}$ as given in step \ref{rowOp:A4} of Algorithm \ref{alg:rowOP}.

\subsubsection*{Remark \ref{subsec:phaseNwSynth}}

In \cite{2020_NGM} an algorithm to synthesize parity networks over $\{\CNOT,R_z\}$ was described and a somewhat similar scheme was sketched in \cite{2019_KdG}. Both used Steiner trees and the sum-over-path formualtion of such circuits. 
Our algorithm is significantly different from both, especially considering the way we assigned CNOT gates according to the constructed Steiner trees. 
An illustration of PHASE-NW-SYNTH has been given in Appendix \ref{app:phase}. 

\section{Synthesis of circuits over $\{\CNOT,\X,\T,\had\}$ gates}
\label{sec:algoH}

Finally in this section we are in a position to describe our re-synthesis algorithms that takes as input a circuit $\ckt_I$ over a universal fault-tolerant gate set $\mathcal{G}_{univ}=\{\CNOT,\T,\T^{\dagger},\phase,\phase^{\dagger},\X,\Y,\Z,\had\}$ and it outputs a circuit $\ckt_O$ with gates in the same set, but the position of the CNOT gates are restricted by some connectivity constraints imposed by an input connectivity graph $G$.

The basic format of our re-synthesis algorithms include \emph{slicing} the given circuit and \emph{building} the sliced portions. We partition the given circuit at the position of the $\had$ gates and then sequentially re-synthesize sub-circuits in each portion such that the transformation within 
each portion and the overall circuit transformation remains unchanged. We investigate two methods of slicing - the first one is a simple \emph{slice-and-build}, where we partition the input circuit according to the position of the H gate and the second one is motivated by the Tpar algorithm given in \cite{2014_AMM}, where the phase polynomial terms are partitioned. Unlike Tpar we are not interested in reducing the T-depth of the input circuit. So we partition the phase polynomial terms depending only on the path variables, which indicate the parities that can be synthesized before or after a certain H gate.  

\subsection{Simply slice-and-build}
\label{subsec:sliceNbuild}
\begin{algorithm}
\footnotesize
 \caption{CNOT-OPT-A}
 \label{alg:slice}
 
 \KwIn{(i) Circuit $\ckt_I$ with $\{\CNOT,\T,\T^{\dagger},\phase,\phase^{\dagger},\X,\Y,\Z,\had\}$ gates, (ii) Connectivity graph $G$.}
 \KwOut{Circuit $\ckt_O$ with $\CNOT$ gates respecting connectivity constraints imposed by $G$. }
 
 $\ckt_O\leftarrow\emptyset$; $\quad i=-1;\quad gate=START$ \;
 
 \While{$gate != EOF$}
 {
    $i\leftarrow i+1;\quad gate=\ckt_O[i]$;$\quad Q=(x_1,x_2,\ldots,x_n); \quad \phpoly\leftarrow\emptyset$  \label{slice:initPQ}\;
    \While{$gate != \had$ or $gate != EOF$ \label{slice:PQbuildStart}}
    {
        Update $Q$ and $\phpoly$ according to the rules described in Section \ref{subsec:sliceNbuild} \;
        $i\leftarrow i+1; \quad gate=\ckt_O[i]$    \;
     \label{slice:PQbuildEnd}
     }
    $\ckt_{ph}, \ckt_{lin}\leftarrow\emptyset$ \;
    $\ckt_{ph}\leftarrow$ PHASE-NW-SYNTH($\phpoly, G$)  \label{slice:Cph}\;
    $Q_{ph}\leftarrow$ state of the qubits after $\ckt_{ph}$   \label{slice:Qph} \;
    $\vect{A}\leftarrow$ linear transformation mapping $Q_{ph}$ to $Q$ \label{slice:A} \;
    $\ckt_{lin}\leftarrow$ LINEAR-TF-SYNTH($\vect{A}, G$)  \label{slice:Clin} \;
    $C_O.append(\ckt_{ph},\ckt_{lin},\had[k])$ \tcp*[f]{$k$ is the position of $\had$} \label{slice:Co} \;
 }
 
 \Return $\mathcal{\ckt}_O$ \;
\end{algorithm}

In our first algorithm CNOT-OPT-A (Algorithm \ref{alg:slice}) we first partition the given circuit at the position of $\had$ gates. Within each partition we initialize the state of the qubits $Q$ by the path variables $(x_1,x_2,\ldots,x_n)$ and the phase polynomial set $\phpoly$ as empty set (step \ref{slice:initPQ}). Then with the application of each gate $U_i\in\mathcal{G}_{univ}$ or $U_{(ij)}\in\mathcal{G}_{univ}$, we update $Q$ and $\phpoly$ (step \ref{slice:PQbuildStart}-step \ref{slice:PQbuildEnd}) by the function $\widetilde{U}_i:\braket{\phpoly,Q}\rightarrow\braket{\phpoly,Q}$ as follows.
\begin{eqnarray}
 &&\widetilde{\X_i}\braket{\phpoly,Q}=\braket{\phpoly,(q_1,\ldots,q_{i-1},1\oplus q_i,\ldots,q_n)} ;   \nonumber \\
 &&\widetilde{\CNOT_{(i,j)}}\braket{\phpoly,Q}=\braket{\phpoly,(q_1,\ldots,q_{j-1},q_i\oplus q_j,\ldots,q_n)} ; \nonumber \\
 &&\widetilde{\phase_i}\braket{\phpoly,Q}=\braket{\phpoly\uplus \{(2,q_i)\},Q};\quad \qquad
 \widetilde{\phase_i^{\dagger}}\braket{\phpoly,Q}=\braket{\phpoly\uplus \{(6,q_i)\},Q} ; \nonumber  \\
 &&\widetilde{\T_i}\braket{\phpoly,Q}=\braket{\phpoly\uplus \{(1,q_i)\},Q};\quad \qquad
 \widetilde{\T_i^{\dagger}}\braket{\phpoly,Q}=\braket{\phpoly\uplus \{(7,q_i)\},Q}; \nonumber \\
 &&\widetilde{\Z_i}\braket{\phpoly,Q}=\braket{\phpoly\uplus \{(4,q_i)\},Q}; \nonumber \\
 &&\widetilde{\Y_i}\braket{\phpoly,Q}=\braket{\phpoly\uplus \{(4,q_i)\},(q_1,\ldots,q_{i-1},1\oplus q_i,\ldots,q_n)}; \nonumber 
\end{eqnarray}
In the above set of equations, for two sets $\phpoly'$ and $\phpoly''$, $\phpoly'\uplus \phpoly''=\{(c,f):(c_1,f)\in \phpoly', (c_2,f)\in \phpoly'', c=c_1+c_2\mod 8 \}$. Here we assume that if $\exists f$ such that $(c,f)\notin \phpoly$ for any $c=\{1,\ldots,7\}$ and any set $\phpoly$, then we say $(0,f)\in\phpoly$. 

Then we synthesize the phase polynomial network ($\ckt_{ph}$) for $\mathcal{P}$ (step \ref{slice:Cph}) by invoking the procedure PHASE-NW-SYNTH (Algorithm \ref{alg:phaseNwSynth}). We calculate the linear transformation $\vect{A}$ (step \ref{slice:A}) mapping $Q_{ph}$ (state of the qubits after $\ckt_{ph}$) to $Q$, which after steps \ref{slice:PQbuildStart}-\ref{slice:PQbuildEnd} stores the state of the qubits at the end of the present slice. We synthesize the circuit $\ckt_{lin}$ for $\vect{A}$ (step \ref{slice:Clin}) using the procedure LINEAR-TF-SYNTH (Algorithm \ref{alg:linTfSynth}). We append the gates from $\ckt_{ph}, \ckt_{lin}$ followed by the $\had$ gate (step \ref{slice:Co}). The we repeat the same steps for the next slice (till the next H gate or the end of the given circuit).

\begin{algorithm}
\footnotesize
 \caption{CNOT-OPT-B}
 \label{alg:cnotOpt}
 
 \KwIn{(i) Circuit $\ckt_I$ with $\{\CNOT,\T,\T^{\dagger},\phase,\phase^{\dagger},\X,\Y,\Z,\had\}$ gates, (ii) Connectivity graph $G$.}
 \KwOut{Circuit $\ckt_O$ with $\CNOT$ gates respecting connectivity constraints imposed by $G$. }
 
 $\ckt_O\leftarrow\emptyset$\;
 From $\ckt_{I}$ calculate $\mathcal{D}=\braket{\phpoly,Q,\hset}$; $\quad k\leftarrow |\hset|$.
 \tcp*[f]{Described in Section \ref{sec:algoH}} \;
 $Q_{init}=(x_1,x_2,\ldots,x_n)$; $\quad Q_{out}=Q$ \;
 \For{$1\leq i\leq k$}
 {
    $h\leftarrow \hset_i$; $\ckt_{ph}\leftarrow\emptyset$; $\ckt_{lin}\leftarrow\emptyset$  \;
    $\phpoly'\leftarrow\{(c,f)\in\phpoly: f \in \Span(h.Q_I) \text{ but } f\notin\Span(h.Q_O)\}$    \label{cnotOPt:P'}  \;
    \If{$\phpoly\neq\emptyset$}
    {
        $\phpoly.delete(\phpoly')$  \;
        $\phpoly_{Qinit}\leftarrow\{(c,f)\in\phpoly' \text{ in the basis } Q_{init}\}$   \label{cnotOpt:Pbasis}\;
        $\ckt_{ph}\leftarrow$PHASE-NW-SYNTH($\phpoly_{Qinit},G$)    \label{cnotOpt:Cph}\;
        $Q_{ph}\leftarrow$ state of the qubits after $\ckt_{ph}$ \label{cnotOpt:Qph}\;
        $\vect{A}\leftarrow$ linear transformation mapping $Q_{ph}$ to $h.Q_I$  \label{cnotOpt:A}\;
        $\ckt_{lin}\leftarrow$LINEAR-TF-SYNTH($\vect{A},G$) \label{cnotOpt:Clin} \;
    }
    $Q_{init}\leftarrow h.Q_O$  \;
    $\ckt_O.append(\ckt_{ph},\ckt_{lin},\had[k])$ where $k=h.Pos$   \; 
 }
 
 $\ckt_{ph}\leftarrow\emptyset$; $\quad\ckt_{lin}\leftarrow\emptyset$ \;
 
 
 \eIf{$\phpoly\neq\emptyset$}
 {
    $\phpoly_{Qinit}=\{(c,f)\in\phpoly \text{ in the basis } Q_{init}\}$    \label{cnotOpt:Pbasis2}\;
    $\ckt_{ph}\leftarrow$PHASE-NW-SYNTH($\phpoly_{Qinit},G$)    \label{cnotOpt:Cph2}\;
    $Q_{ph}\leftarrow$ state of the qubits after $\ckt_{ph}$     \label{cnotOpt:Qph2}\;
 }
 {
    $Q_{ph}\leftarrow Q_{init}$  \;
 }
 $\vect{A}\leftarrow$ linear transformation mapping $Q_{ph}$ to $Q_{out}$ \label{cnotOpt:A2}\;
 $\ckt_{lin}\leftarrow$LINEAR-TF-SYNTH($\vect{A},G$) \label{cnotOpt:Clin2}  \;
 $\ckt_O.append(\ckt_{ph},\ckt_{lin})$  \; 
 
 \Return $\mathcal{\ckt}_O$ \;
\end{algorithm}
\begin{table*}
\centering
\scriptsize
\begin{tabular}{|c|c|c|c|c|c|c|c|c|c|}
 \hline
 Architecture & $\#$Qubits & Initial & SWAP-template & \multicolumn{2}{|c|}{CNOT-OPT-A} & \multicolumn{2}{|c|}{CNOT-OPT-B} & Qiskit & TKET\\
 \cline{5-8}
 & & Count & Overhead & Overhead & Time & Overhead & Time & Overhead & Overhead\\
 \hline\hline
 \multirow{5}{*}{9q-square} & \multirow{5}{*}{9} & 3 & $560\%$ & $0.00\%$& 0.184s& $343\%$ & 0.105s & $310\%$ & $180\%$\\
 \cline{3-10}
 & & 5 & $612\%$ &$146\%$ &0.146s & $400\%$ & 0.128s & $300\%$ & $218\%$\\
 \cline{3-10}
 & & 10 & $594\%$ & $105\%$& 0.167s & $426\%$ & 0.119s & $303\%$ & $177\%$\\
 \cline{3-10}
 & & 20 & $546\%$ &$176\%$ & 0.2s & $488\%$ & 0.158s & $265.50\%$ & $154\%$\\
 \cline{3-10}
 & & 30 & $596\%$ & $184.67\%$& 0.233s & $649\%$ & 0.185s & $310\%$ & $165.33\%$\\
 \hline\hline
 \multirow{7}{*}{16q-square} & \multirow{7}{*}{16} & 4 & $1050\%$ & $238\%$& 0.23s & $768\%$ & 0.12s & $562.50\%$ & $465\%$\\
 \cline{3-10}
 & & 8 & $840\%$ &$146.25\%$ & 0.27s & $660\%$ & 0.137s & $461.25\%$ & $321.25\%$\\
 \cline{3-10}
 & & 16 & $817.50\%$ & $158.13\%$& 0.43s & $864\%$ & 0.225s & $470.63\%$ & $306.88\%$\\
 \cline{3-10}
 & & 32 & $853\%$ & $340.63\%$& 0.41s & $1213\%$ & 0.29s & $493.13\%$ & $309.38\%$\\
 \cline{3-10}
 & & 64 & $892.50\%$ & $220.78\%$& 0.49s & $1259\%$ & 0.65s & $476.25\%$ & $276.09\%$\\
 \cline{3-10}
 & & 128 & $858.75\%$ & $210.63\%$& 0.57s& $1156\%$ & 1.144s & $475.31\%$ & $276.8\%$\\
 \cline{3-10}
 & & 256 & $897.42\%$ &$237.5\%$ & 0.72s & $1306\%$ & 1.85s & $497.46\%$ & $276.6\%$\\
 \hline\hline
 \multirow{7}{*}{rigetti-16q-aspen} & \multirow{7}{*}{16} & 4 & $1680\%$ &$355\%$ &0.23s & $1278\%$ & 0.115s & $750\%$ & $645\%$\\
 \cline{3-10}
 & & 8 & $1740\%$ & $253\%$& 0.396s& $1313\%$ & 0.135s & $738.75\%$ & $578.75\%$\\
 \cline{3-10}
 & & 16 & $1619.90\%$ & $351\%$& 0.47s& $1304\%$ & 0.162s & $701.25\%$ & $454.38\%$\\
 \cline{3-10}
 & & 32 & $1794\%$ & $469.48\%$& 0.48s & $1852\%$ & 0.375s & $732.19\%$ & $495.63\%$\\
 \cline{3-10}
 & & 64 & $1755\%$ &$399\%$ & 0.66s & $1900\%$ & 0.71s & $728.91\%$ & $476.25\%$\\
 \cline{3-10}
 & & 128 & $1760.63\%$ & $368.13\%$& 0.58s & $1953\%$ & 1.37s & $731.02\%$ & $471.17\%$\\
 \cline{3-10}
 & & 256 & $1757.11\%$ &$410.9\%$ & 0.61s & $1982\%$ & 1.68s & $725.74\%$ & $463.44\%$\\
 \hline\hline
 \multirow{7}{*}{ibm-qx5} & \multirow{7}{*}{16} & 4 & $1260\%$ &$173\%$ & 0.38s & $988\%$ & 0.108s & $690\%$ & $547.5\%$\\
 \cline{3-10}
 & & 8 & $1035\%$ &$295\%$ & 0.36s & $1065\%$ & 0.126s & $712.50\%$ & $517.5\%$\\
 \cline{3-10}
 & & 16 & $1042.50\%$ & $283\%$& 0.41s &$1226\%$ & 0.47s & $710.63\%$ & $469.38\%$\\
 \cline{3-10}
 & & 32 & $1179.38\%$ &$398.44\%$ & 0.42s& $1677\%$& 0.68s & $670.31\%$ & $447.5\%$\\
 \cline{3-10}
 & & 64 & $1130.63\%$ &$339.06\%$ &0.45s & $1733\%$& 0.7s & $646.88\%$ & $441.56\%$\\
 \cline{3-10}
 & & 128 & $1110.94\%$ &$344.69\%$ &0.575s & $1675\%$ &1.15s & $689.30\%$ & $434.45\%$\\
 \cline{3-10}
 & & 256 & $1141.17\%$ & $379.88\%$ &0.73s & $1792\%$ &1.58s & $678.52\%$ & $434.1\%$\\
 \hline\hline
 \multirow{7}{*}{ibm-q20-tokyo} & \multirow{7}{*}{20} & 4 & $525\%$ & $128\%$ & 0.186s & $418\%$ & 0.4s & $435\%$ & $315\%$\\
 \cline{3-10}
 & & 8 & $555\%$ &$275\%$ &0.295s & $690\%$&0.37s & $506.25\%$ & $293.75\%$\\
 \cline{3-10}
 & & 16 & $570\%$ &$88\%$ & 0.37s& $663\%$&0.41s & $472.50\%$ & $257.5\%$\\
 \cline{3-10}
 & & 32 & $500.63\%$ &$154.38\%$ & 0.55s& $972\%$&0.8s & $420\%$ & $235.63\%$\\
 \cline{3-10}
 & & 64 & $542.81\%$ &$136.88\%$ &0.54s & $1084\%$ &0.82s & $463.59\%$ & $229.22\%$\\
 \cline{3-10}
 & & 128 & $539.53\%$ &$141.02\%$ & 0.645s&$1028\%$ &1.29s & $468.05\%$ & $228.52\%$\\
 \cline{3-10}
 & & 256 & $534.61\%$ &$125.27\%$ & 0.72s & $1030\%$ & 2.085s & $457.50\%$ & $226.37\%$\\
 \hline
\end{tabular}
\caption{Performance of CNOT-OPT-A and CNOT-OPT-B for random circuits. The overhead i.e. $\frac{\text{Final count - Initial count}}{\text{Initial count}}\times 100\%$ has been compared to the overhead obtained by using SWAP-template, Qiskit and TKET transpiler.}
\label{tab:random}
\end{table*}

\begin{table*}
\centering
\scriptsize
 \begin{tabular}{|c|c|c|c|c|c|c|c|c|c|c|}
  \hline
  Architecture & $\#$Qubits & Benchmark & SWAP-template & \multicolumn{2}{|c|}{CNOT-OPT-A} & \multicolumn{2}{|c|}{CNOT-OPT-B} & Qiskit & TKET\\
 \cline{5-8}
 & & & Overhead & Overhead & Time & Overhead & Time & Overhead & Overhead\\
 \hline\hline
 \multirow{4}{*}{9q-square} & \multirow{4}{*}{9} & barenco-tof-5 & $457.14\%$ &$245.24\%$ &0.365s & $140.48\%$ & 0.52s & $232.14\%$ & $110.71\%$\\
 \cline{3-10}
 & & grover-5 & $685.71\%$ &$116.67\%$ &0.502s & $105.36\%$ & 0.84s & $219.64\%$ & $94.05\%$\\
 \cline{3-10}
 & & mod-mult-55 & $752.73\%$ &$321.82\%$ &0.31s & $203.64\%$ & 0.26s & $256.36\%$ & $147.27\%$\\
 \cline{3-10}
 & & tof-5 & $465.31\%$ &$140.82\%$ & 0.27s & $138.78\%$ & 0.24s & $226.53\%$ & $110.20\%$\\
 \hline\hline
 16q-square & 16 & \multirow{3}{*}{hwb10} & $977.95\%$ &$-63\%$ & 8.77s & $-57.67\%$ & 7.25s & $284.76\%$ & $133.10\%$\\
 \cline{1-2}\cline{4-10}
 rigetti-16q-aspen & 16 & & $1508.63\%$ &$-36.13\%$ &8.8s & $-34.29\%$ & 6.64s & $356.70\%$ & $187.02\%$\\
 \cline{1-2}\cline{4-10}
 ibm-qx5 & 16 & & $1099.84\%$ & $-54.32\%$& 6.18s & $-50.75\%$ & 8.61s & $321.71\%$ & $160.32\%$\\
 \hline\hline
 \multirow{2}{*}{ibm-q20-tokyo} & \multirow{2}{*}{20} & ham15-high & $571.44\%$ & $-52.52\%$& 0.72s & $-63.21\%$ & 0.66s & $242.76\%$ & $72.92\%$\\
 \cline{3-10}
 & & hwb12 & $619.42\%$ & $-77.58\%$ & 177.23s & $-74.92\%$ & 231.14s & $277.82\%$ & $87.81\%$\\
 \hline
 \end{tabular}
\caption{Performance of CNOT-OPT-A and CNOT-OPT-B for benchmark circuits. The overhead has been compared to the overhead obtained by using SWAP-template, Qiskit and TKET transpiler.}
\label{tab:benchmark}
\end{table*}

\subsection{A second type of slice-and-build}
\label{subsec:sliceNbuild2}

In this section we describe another way of slicing the given circuit, as described in procedure CNOT-OPT-B (Algorithm \ref{alg:cnotOpt}). Unlike CNOT-OPT-A, here we first compute some necessary information about the whole circuit and then between two H gates we try to synthesize a circuit that computes part of the information. Similar to CNOT-OPT-A the transformations between two H gates as well as the overall transformation remain unchanged. That is, given $\ckt_I$, we first compute a triple $\mathcal{D}=\braket{\mathcal{P},Q,\hset}$, where $\mathcal{P}$ is the phase polynomial set, $Q=\left(q_1,q_2,\ldots,q_n\right)$ represents the state of each qubit given as a function of the path variables and the bit flip variable $b\in\{0,1\}$, and $\hset$ is an array of structures where the $i^{th}$ structure stores the state of the qubits before and after the application of the $i^{th}$ $\had$ gate. The initial state of the qubits is $Q=\left(x_1,x_2,\ldots,x_n \right)$ (i.e. $q_i=x_i\quad\forall i$). Both $\mathcal{P}$ and $\hset$ are initialized as empty sets. With the application of each gate $U_i\in\mathcal{G}_{univ}$ or $U_{(i,j)}\in\mathcal{G}_{univ}$ (subscripts denote the qubit on which the gate acts) the triple $\mathcal{D}$ gets updated by a function $\widetilde{U_i'}:\mathcal{D}\rightarrow\mathcal{D}$. Except for the H gate, this function is similar to  the function $\widetilde{U}_i$ defined in Section \ref{subsec:sliceNbuild}. The array $\hset$ remains unchanged after the application of $\widetilde{U_i'}$ for each gate except H. For the H gate the function is defined as follows.
\begin{eqnarray}
 &&\widetilde{\had_i} \braket{\phpoly,Q,\hset}=\braket{\phpoly,Q',\hset'} \text{ where } Q'=(q_1,\ldots,q_{i-1},x_{n+j+1},\ldots,q_n) \text{ and } \hset'=\hset\cup\{h_{j+1}\} \nonumber \\
 &&\text{ such that } h_{j+1}.Pos=i,\quad h_{j+1}.Q_I=Q,\quad h_{j+1}.Q_O=Q'
 \quad [\text{ Here } |\hset|=j.] \nonumber 
\end{eqnarray}
$h_{j+1}.Pos$ stores the qubit (which in the above equation is the $i^{th}$ one) on which the $(j+1)^{th}$ H gate is applied. $h_{j+1}.Q_I$ and $h_{j+1}.Q_O$ stores the state of the qubits before and after the application of the $(j+1)^{th}$ H gate respectively.
We have introduced new path variables after application of each H gate. We actually slice the sets $\phpoly, Q$ and $\hset$ according to some conditions and synthesize circuits according to these slices.

For each $h\in \mathcal{H}$ we calculate the set $\mathcal{P}'=\{(c,f)\in\phpoly:f\in\Span(h.Q_I) \text{ but } f\notin\Span(h.Q_O)\}\subseteq\phpoly$ of parity terms that become uncomputable after placement of $\had$ at qubit $h.Pos$ (step \ref{cnotOPt:P'} in Algorithm \ref{alg:cnotOpt}). We express these parities in the basis given by the state of the qubits at the beginning of the current time slice, which is $Q_{init}=(x_1,\ldots,x_n)$ if $h$ is the first Hadamard gate, else it is $h'.Q_O$, the state of the qubits after the previous $\had$ gate (step \ref{cnotOpt:Pbasis}). We then calculate the phase polynomial network ($\ckt_{ph}$) for the set $\phpoly_{Qinit}$ ($\phpoly'$ in the new basis i.e. parity terms in $\phpoly'$ as function of $Q_{init}$) by invoking the procedure PHASE-NW-SYNTH (Algorithm \ref{alg:phaseNwSynth}). Let $Q_{ph}$ is the state of the qubits after $\ckt_{ph}$ and $\vect{A}$ is the linear transformation mapping $Q_{ph}$ to $h.Q_I$. To realize this transformation in this portion of the circuit we call the procedure LINEAR-TF-SYNTH (Algorithm \ref{alg:linTfSynth}). We append ($\ckt_{ph}$, $\ckt_{lin}$,$H[k]$) to the set of circuit gates, where $\ckt_{lin}$ is the circuit returned by LINEAR-TF-SYNTH and $k=h.Pos$.

After we process all the partitions till the last Hadamard gate we ensure that the complete phase polynomial set $\phpoly$ has been synthesized and the overall linear transformation of the output circuit maps $(x_1,x_2,\ldots,x_n)$ to $Q_{out}$, the final output of the circuit (which was calculated at the beginning while calculating $\mathcal{D}$). For this we first synthesize the phase polynomial network $\ckt_{ph}$ of any residual parity terms (step \ref{cnotOpt:Cph2} in Algorithm \ref{alg:cnotOpt}). Then we calculate the residual transformation $\vect{A}$ (step \ref{cnotOpt:A2}) that maps $Q_{ph}$, state of the qubits after $\ckt_{ph}$, to $Q_{out}$. And synthesize the circuit $\ckt_{lin}$ (step \ref{cnotOpt:Clin2}) for $\vect{A}$.

\subsection{Implementation and results}
\label{subsec:result}

We have considered the connectivity constraints imposed by some popular architectures such as 9-qubit square grid (Figure \ref{fig:swapTemplate}), 16-qubit square grid, Rigetti 16-qubit Aspen, 16-qubit IBM QX5 and 20-qubit IBM Tokyo (Figure \ref{fig:arch}). We have worked with some benchmark circuits (Table \ref{tab:benchmark}) and some randomly generated circuits on 9, 16 and 20 qubits (Figure \ref{tab:random}). The 9-qubit random input circuits have CNOT-count 3, 5, 10, 20 or 30, while both the 16 and 20-qubit random input circuits have CNOT-count 4, 8, 16, 32, 64, 128 or 256. For each of these groups we generated 10 random circuits. We have compared the CNOT-count overhead obtained by using SWAP-template (Figure \ref{fig:swapTemplate}), IBM's Qiskit \cite{2021_Qiskit} and TKET \cite{2021_SDCetal} transpiler with the CNOT-count obtained from procedures CNOT-OPT-A (Algorithm \ref{alg:slice}) and CNOT-OPT-B (Algorithm \ref{alg:cnotOpt}). By overhead we mean the percentage increase in CNOT-count after taking into consideration connectivity constraints i.e. $\frac{\text{Final count - Initial count}}{\text{Initial count}}\times 100\%$. The results for benchmark circuits and the random circuits have been tabulated in Table \ref{tab:benchmark} and \ref{tab:random} respectively. All the simulations have been done in Java on a 3.1 GHz Dual-Core Intel Core i7 machine with 8GB RAM and running MacOS Catalina 10.15.2. In Qiskit we ran each circuit with optimization level till 4 and then took the average. The ``layout-method'' is 'trivial' (no qubit remapping).
We find that both our algorithms perform quite well in the case of benchmark circuits. CNOT-OPT-A performs much better than the other algorithms in the case of random circuits. In CNOT-OPT-B we compute the phase polynomial of entire circuit and then do slicing. If we re-synthesize a circuit from its phase polynomial then usually the gates which change the phase, like T-gate, gets reduced. This comes at the cost of increase in other gates like CNOT. Such observations have also been made in \cite{2014_AMM}, though they do not consider connectivity constraints. In contrast, CNOT-OPT-A slices the entire circuit and then computes the phase polynomial of the segments. Although there is still some optimization of T-gates that may increase the CNOT-count, but this time it happens in a ``more localized scale''. 
This can be one reason for a less significant overall increment of CNOTs, compared to CNOT-OPT-B. In fact, we think that any algorithm that synthesizes  or re-synthesizes using phase terms will tend to favour the reduction of gates like T. Hence, in CNOT-OPT-A this effect is mitigated to some extent by confining this reduction in shorter segments, leading to a better performance compared to CNOT-OPT-B.

Software packages like Qiskit and TKET have other built-in features like efficient initial qubit mapping, that aims to improve their performance. We have considered only one mapping, that is, logical qubit i is mapped to vertex i in the connectivity graph. If we change this mapping, it may be possible to get even better results. Hence, it is not completely fair for us to compare our results with these software packages. Despite such difference, our CNOT-OPT-A algorithm gives much less overhead compared to Qiskit and TKET in most cases. 

\section{Conclusion}
\label{sec:conclude}

While implementing a quantum algorithm on an actual hardware, one needs to consider the different constraints imposed by the underlying physical architecture. One such constraint is the qubit connectivity, which is more concerning for multi-qubit gates such as CNOT. 
In a universal fault-tolerant gate set such as CNOT+T, although the T-gate is the most costly to implement fault-tolerantly, the CNOT-count is also important, especially in the NISQ era. In this work we have considered the problem of re-synthesizing Clifford+T circuits with reduced CNOT-count compared to the SWAP-template, while respecting the connectivity constraint.
Broadly, we have taken recourse to a slice-and-build approach, where we slice or partition the input circuit and re-synthesize the slices with algorithms that use Steiner trees to place the CNOT gates. 
We have simulated benchmarks as well as random circuits on popular architectures such as 9-qubit square grid, 16-qubit square grid, Rigetti 16-qubit Aspen, 16-qubit IBM QX5 and 20-qubit IBM Tokyo. Our results show that for both benchmarks and random circuits the simpler way of slicing the circuit (and not the phase polynomial) results in much less overhead in terms of increase in CNOT-count, compared to the overhead obtained by using SWAP-template, Qiskit and TKET transpiler.

\section*{Acknowledgement}

This work was supported in part by Canada's NSERC.  IQC and the Perimeter Institute (PI) are supported in part by the Government of Canada and Province of Ontario (PI). A significant part of this work was done while S.M.Li was a student at the Department of Mathematics and Statistics, Dalhousie University, Halifax NS, Canada and was an under-graduate research intern at IQC.

\section*{Author contributions}

The ideas were given by P.Mukhopadhyay. The implementations were done by S.M.Li and J.Huang. All the authors contributed in preparation of the manuscript.

\section*{Data availability}

The results obtained in this paper are available in a Github repository at \\
\url{https://github.com/SarahMLi/cnotcount}.

\section*{Code availability}

The code is available from the corresponding author on reasonable request.

\section*{Competing interests}
Michele Mosca is co-founder of softwareQ Inc. and has filed a provisional patent application for this work. P.Mukhopadhyay and V.Gheorghiu are co-inventors.

 \newcommand{\etalchar}[1]{$^{#1}$}

\appendix
\section{Steiner tree algorithm}
\label{app:steinerTree}

In this section we describe a heuristic approximation algorithm to find a minimal Steiner tree \cite{2013_SF}. It starts by considering each terminal as a separate graph (step \ref{stTree:fi}). Then sequentially we merge the subgraphs that are closest to each other (step \ref{stTree:shortestPath}-\ref{stTree:merge}). The distance between two graphs $f_i$ and $f_j$ is measured by the shortest distance between any pair of nodes $u_i,u_j$ such that $u_i\in f_i$ and $u_j\in f_j$ (step \ref{stTree:distNode}). If we have a sub-graph $f_i$ having all the terminal nodes then we construct a minimum spanning tree (\ref{stTree:mst}) $T$ on $f_i$ and remove all non-terminal nodes of degree 1 \ref{stTree:nonTerm}. The resultant tree is returned as a minimum Steiner tree. 

\begin{algorithm}
 \caption{Steiner Tree Algorithm}
 \label{alg:steinerTree}
 
 \KwIn{(i) A graph $G=(V,E)$, (ii) Terminal set $\tset=\{s_1,\ldots,s_k\}\subseteq V$}
 \KwOut{A Steiner tree constructed from $G$}
 
 Construct a forest $F$ of $k$ sub-graphs $f_1,\ldots,f_k$ consisting of one terminal each \label{stTree:fi}\;
 \While{does not exist a $f_i\in F$ such that all terminals $s_1,\ldots,s_k\inf_i$}
 {
    \For{all $i\neq j$ \label{stTree:shortestPath}}
    {
        Determine the shortest path between all nodes in $f_i$ to all those in $f_j$ \;
        \label{stTree:distNode}
    }
    Find the minimum length path $P$ among all computed paths \;
    Construct $f_n=f_i\cap f_j \cap P$ and add it to forest $F$ \label{stTree:merge}\;
 }
 Construct a minimum spanning tree $T$ on $f_i$  \label{stTree:mst}\;
 Remove non-terminal nodes of degree $1$ from $T$  \label{stTree:nonTerm} \;
 \Return $T$    \;
\end{algorithm}

\section{An example for LINEAR-TF-SYNTH}
\label{app:lin}
        
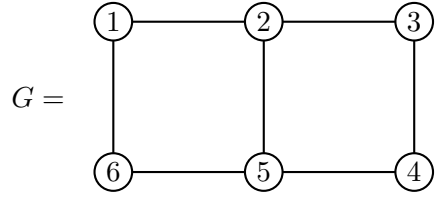
\begin{figure}[h]
\centering
 \begin{tikzpicture}  
 $A= \begin{bmatrix}
     1 & 1 & 0 & 1 & 1 & 0 \\
     0 & 0 & 1 & 1 & 0 & 1 \\
     1 & 0 & 1 & 0 & 1 & 0 \\
     1 & 1 & 0 & 1 & 0 & 0 \\
     1 & 1 & 1 & 1 & 0 & 0 \\
     0 & 1 & 0 & 1 & 0 & 1
    \end{bmatrix}
$

\node at (3,0) {$G=$};
  \draw [thick] (4,1) circle [radius=0.25];
  \node at (4,1) {$1$};
  \draw [thick] (4.25,1)--(5.75,1); 
  \draw [thick] (4.25,-1)--(5.75,-1); 
  \draw [thick] (6,1) circle [radius=0.25];
  \node at (6,1) {$2$};
  \draw [thick] (6.25,1)--(7.75,1); 
  \draw [thick] (6.25,-1)--(7.75,-1); 
  \draw [thick] (8,1) circle [radius=0.25];
  \node at (8,1) {$3$};
  \draw [thick] (4,-1) circle [radius=0.25];
  \node at (4,-1) {$6$};
  \draw [thick] (6,-1) circle [radius=0.25];
  \node at (6,-1) {$5$};
  \draw [thick] (8,-1) circle [radius=0.25];
  \node at (8,-1) {$4$};
  \draw [thick] (4,0.75)--(4,-0.75);    
  \draw [thick] (8,0.75)--(8,-0.75);    
  \draw [thick] (6,0.75)--(6,-0.75); 
 \end{tikzpicture}
\caption{The linear transformation matrix $A$ and connectivity graph $G$.}
\label{fig:app:lin0}
\end{figure}

In this section we illustrate Algorithm \ref{alg:linTfSynth} (LINEAR-TF-SYNTH) with an example taken from \cite{2020_NGM}. Suppose we have the linear tranformation matrix $A$ and connecitvity graph $G$, as shown in Figure \ref{fig:app:lin0}. For simplicity, we have removed the column $\vect{b}$ ($7^{th}$ column of $A$), since it is all 0 and hence implies there are no $\X$ gates to be placed. So this column does not affect any step of the following computations, which are done for the placement of the CNOT gates.

\subsection{Reducing to upper triangular form}

We now illustrate the steps to reduce $A$ to upper triangular form and the way the CNOTs are placed respecting the connecitvity constraints. For each column $i$, the pivot row is the $i^{th}$ one. By ``fixing'' a column $i$ we mean applying a number of row operations such that $A_{ii}=1$ and $A_{ji}=0$ for every $j>i$. We start fixing from the first column. To save redundancies in explanations, we will make the descriptions from the second column less explicit.

\subsection*{Column 1}

In this column $A_{11}=A_{31}=A_{41}=A_{51}=1$. So we draw a Steiner tree $T_{1,\mathcal{S}}$ with terminals $\mathcal{S}=\{1,3,4,5\}$ and root/pivot at $1$ (Figure \ref{fig:app:linUp1}). We call this ``pivot'', to distinguish from the roots of each sub-tree  that will be built from this tree. The sub-trees $T_1, T_2$ and $T_3$ are built by traversing $T_{1,\mathcal{S}}$ breadth-first (Algorithm \ref{alg:separate}:SEPARATE). As soon as we reach a terminal we stop and build the next sub-tree with that terminal as the root. Thus in each sub-tree, the root and leaves are terminals and remaining nodes are Steiner nodes.

\begin{figure}
\centering
\begin{subfigure}{0.3\textwidth}
\centering
\footnotesize
$A= \begin{bmatrix}
     \mathbf{1} & 1 & 0 & 1 & 1 & 0 \\
     \mathbf{0} & 0 & 1 & 1 & 0 & 1 \\
     \mathbf{1} & 0 & 1 & 0 & 1 & 0 \\
     \mathbf{1} & 1 & 0 & 1 & 0 & 0 \\
     \mathbf{1} & 1 & 1 & 1 & 0 & 0 \\
     \mathbf{0} & 1 & 0 & 1 & 0 & 1
    \end{bmatrix}
$
\end{subfigure}
\hfill
\begin{subfigure}{0.65\textwidth}
\centering
\footnotesize
\begin{tikzpicture}
\node at (-0.5,0.5) {$T_{1,\{1,3,4,5\}}$=};
 \draw [fill=black] (1,1) circle [radius=0.2];
  \node [white] at (1,1) {$1$};
  \draw [ultra thick] (1.2,1)--(1.8,1); 
  \draw (1.2,0)--(1.8,0); 
  \draw (2,1) circle [radius=0.2];
  \node at (2,1) {$2$};
  \draw [ultra thick] (2.2,1)--(2.8,1); 
  \draw [ultra thick] (2.2,0)--(2.8,0); 
  \draw [ultra thick] (3,1) circle [radius=0.2];
  \node at (3,1) {$3$};
  \draw (1,0) circle [radius=0.2];
  \node at (1,0) {$6$};
  \draw [ultra thick] (2,0) circle [radius=0.2];
  \node at (2,0) {$5$};
  \draw [ultra thick] (3,0) circle [radius=0.2];
  \node at (3,0) {$4$};
  \draw (1,0.8)--(1,0.2);    
  \draw [ultra thick] (3,0.8)--(3,0.2);    
  \draw (2,0.8)--(2,0.2); 
  
  \node at (4.5,1.5) {$T_1=$};
  \draw [fill=black] (5.5,1.5) circle [radius=0.2];
  \node [white] at (5.5,1.5) {$1$};
  \draw [ultra thick] (5.7,1.5)--(6.3,1.5);
  \draw (6.5,1.5) circle [radius = 0.2];
  \node at (6.5,1.5) {$2$};
  \draw [ultra thick] (6.7,1.5)--(7.3,1.5);
  \draw [ultra thick] (7.5,1.5) circle [radius=0.2];
  \node at (7.5,1.5) {$3$};
  
  \node at (4.5,0.5) {$T_2=$};
  \draw [ultra thick] (5.5,0.5) circle [radius=0.2];
  \node at (5.5,0.5) {$3$};
  \draw [ultra thick] (5.7,0.5)--(6.3,0.5);
  \draw [ultra thick] (6.5,0.5) circle [radius=0.2];
  \node at (6.5,0.5) {$4$};
  
  \node at (4.5,-0.5) {$T_3=$};
  \draw [ultra thick] (5.5,-0.5) circle [radius=0.2];
  \node at (5.5,-0.5) {$4$};
  \draw [ultra thick] (5.7,-0.5)--(6.3,-0.5);
  \draw [ultra thick] (6.5,-0.5) circle [radius=0.2];
  \node at (6.5,-0.5) {$5$};

\end{tikzpicture}
\end{subfigure}
\caption{The Steiner tree $T_{1,\mathcal{S}}$ with pivot at 1 and terminals $\mathcal{S}=\{1,3,4,5\}$. $T_1, T_2$ and $T_3$ are the sub-trees built from it.}
\label{fig:app:linUp1}
\end{figure}

To place the CNOTs we start from the last sub-tree $T_3$ and use Algorithm \ref{alg:rowOP} (ROW-OP). We have to perform Top-Down-1 first and place $\CNOT_{45}$. The $4^{th}$ row of $A$ gets XORed with the $5^{th}$ row of $A$. If we denote the $j^{th}$ row of $A$ by $A[j,.]$ then $A[5,.]\leftarrow A[5,.]\oplus A[4,.]$, remaining rows are unchanged. There are no operations for Bottom-Up-2 in case of $T_3$, since there are no Steiner nodes. By similar reasoning we traverse by Top-Down-1 in $T_2$ and place $\CNOT_{34}$, while $A[4,.]\leftarrow A[4,.]\oplus A[3,.]$. There are no operations for Bottom-Up-2 in case of $T_2$ as well. We apply Top-Down-1 in $T_1$ and get the sequence of operations $\CNOT_{12},\CNOT_{23}$ and the corresponding sequence of matrix operations $A[2,.]\leftarrow A[2,.]\oplus A[1,.]$ and $A[3,.]\leftarrow A[3,.]\oplus A[2,.]$. Then we apply Bottom-Up-2 and place $\CNOT_{12}$. The corresponding matrix operations are $A[2,.]\leftarrow A[2,.]\oplus A[1,.]$.

Thus the sequence of CNOT operations obtained and the corresponding row operations can be summarized as follows:
\begin{eqnarray}
 \mathcal{Y}_1^1&=&\left\{\CNOT_{45},\CNOT_{34},\CNOT_{12},\CNOT_{23},\CNOT_{12}\right\}   \nonumber \\
 \mathcal{A}_1^1&=&\{A[5,.]\leftarrow A[5,.]\oplus A[4,.], A[4,.]\leftarrow A[4,.]\oplus A[3,.], A[2,.]\leftarrow A[2,.]\oplus A[1,.],  \nonumber \\
 && A[3,.]\leftarrow A[3,.]\oplus A[2,.],A[2,.]\leftarrow A[2,.]\oplus A[1,.]\}    \nonumber
\end{eqnarray}
The matrix $A$ after all these operations is as follows :
$
 A=\begin{bmatrix}
  1 & 1 & 0 & 1 & 1 & 0 \\
  0 & 0 & 1 & 1 & 0 & 1 \\
  0 & 1 & 0 & 0 & 0 & 1 \\
  0 & 1 & 1 & 1 & 1 & 0 \\
  0 & 0 & 1 & 0 & 0 & 0 \\
  0 & 1 & 0 & 1 & 0 & 1
 \end{bmatrix}
$

\subsection*{Column 2}

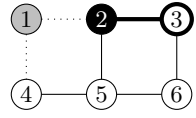
\begin{figure}
\centering
\begin{subfigure}{0.3\textwidth}
\centering
 \footnotesize
$A= \begin{bmatrix}
     1 & \mathbf{1} & 0 & 1 & 1 & 0 \\
     0 & \mathbf{0} & 1 & 1 & 0 & 1 \\
     0 & \mathbf{1} & 0 & 0 & 0 & 1 \\
     0 & \mathbf{1} & 1 & 1 & 1 & 0 \\
     0 & \mathbf{0} & 1 & 0 & 0 & 0 \\
     0 & \mathbf{1} & 0 & 1 & 0 & 1
    \end{bmatrix}
$
\end{subfigure}
\hfill
\begin{subfigure}{0.65\textwidth}
 \centering
 \footnotesize
\begin{tikzpicture}
 \draw [fill=lightgray] (1,1) circle [radius=0.2];
  \node at (1,1) {$1$};
  \draw [dotted] (1.2,1)--(1.8,1); 
  \draw (1.2,0)--(1.8,0); 
  \draw [fill=black] (2,1) circle [radius=0.2];
  \node [white] at (2,1) {$2$};
  \draw [ultra thick] (2.2,1)--(2.8,1); 
  \draw (2.2,0)--(2.8,0); 
  \draw [ultra thick] (3,1) circle [radius=0.2];
  \node at (3,1) {$3$};
  \draw (3,0) circle [radius=0.2];
  \node at (3,0) {$6$};
  \draw (2,0) circle [radius=0.2];
  \node at (2,0) {$5$};
  \draw (1,0) circle [radius=0.2];
  \node at (1,0) {$4$};
  \draw [dotted] (1,0.8)--(1,0.2);    
  \draw (2,0.8)--(2,0.2);    
  \draw (3,0.8)--(3,0.2); 
  
\end{tikzpicture}
\end{subfigure}
\caption{Since $A_{22}=0$ so we apply $\CNOT_{32}$ and ``propagate'' 1 from $A_{32}$ to $A_{22}$.}
\label{fig:app:linUp2_0} 
\end{figure}

\begin{figure}
\centering
\begin{subfigure}{0.3\textwidth}
\centering
 \footnotesize
$A= \begin{bmatrix}
     1 & \mathbf{1} & 0 & 1 & 1 & 0 \\
     0 & \mathbf{1} & 1 & 1 & 0 & 0 \\
     0 & \mathbf{1} & 0 & 0 & 0 & 1 \\
     0 & \mathbf{1} & 1 & 1 & 1 & 0 \\
     0 & \mathbf{0} & 1 & 0 & 0 & 0 \\
     0 & \mathbf{1} & 0 & 1 & 0 & 1
    \end{bmatrix}
$
\end{subfigure}
\hfill 
\begin{subfigure}{0.65\textwidth}
 \centering
 \footnotesize
\begin{tikzpicture}
\node at (-0.5,0.5) {$T_{2,\{2,3,4,6\}}$=};
 \draw [fill=lightgray] (1,1) circle [radius=0.2];
  \node at (1,1) {$1$};
  \draw [dotted] (1.2,1)--(1.8,1); 
  \draw [ultra thick] (1.2,0)--(1.8,0); 
  \draw [fill=black] (2,1) circle [radius=0.2];
  \node [white] at (2,1) {$2$};
  \draw [ultra thick] (2.2,1)--(2.8,1); 
  \draw (2.2,0)--(2.8,0); 
  \draw [ultra thick] (3,1) circle [radius=0.2];
  \node at (3,1) {$3$};
  \draw [ultra thick] (1,0) circle [radius=0.2];
  \node at (1,0) {$6$};
  \draw (2,0) circle [radius=0.2];
  \node at (2,0) {$5$};
  \draw [ultra thick] (3,0) circle [radius=0.2];
  \node at (3,0) {$4$};
  \draw [dotted] (1,0.8)--(1,0.2);    
  \draw [ultra thick] (2,0.8)--(2,0.2);    
  \draw [ultra thick] (3,0.8)--(3,0.2); 
  
  \node at (4.5,1.5) {$T_1=$};
  \draw [fill=black] (5.5,1.5) circle [radius=0.2];
  \node [white] at (5.5,1.5) {$2$};
  \draw [ultra thick] (5.7,1.5)--(6.3,1.5);
  \draw (6.5,1.5) circle [radius = 0.2];
  \node at (6.5,1.5) {$5$};
  \draw [ultra thick] (6.7,1.5)--(7.3,1.5);
  \draw [ultra thick] (7.5,1.5) circle [radius=0.2];
  \node at (7.5,1.5) {$6$};
  \draw [ultra thick] (5.5,0.5) circle [radius=0.2];
  \node at (5.5,0.5) {$3$};
  \draw [ultra thick] (5.5,1.3)--(5.5,0.7);
  
  \node at (4.5,-0.5) {$T_2=$};
  \draw [ultra thick] (5.5,-0.5) circle [radius=0.2];
  \node at (5.5,-0.5) {$3$};
  \draw [ultra thick] (5.7,-0.5)--(6.3,-0.5);
  \draw [ultra thick] (6.5,-0.5) circle [radius=0.2];
  \node at (6.5,-0.5) {$4$};
  
\end{tikzpicture}
\end{subfigure}
\caption{The Steiner tree $T_{2,\mathcal{S}}$ with pivot at 2 and terminals $\mathcal{S}=\{2,3,4,6\}$. $T_1$ and $T_2$ are the sub-trees built from it.}
\label{fig:app:linUp2}
\end{figure}
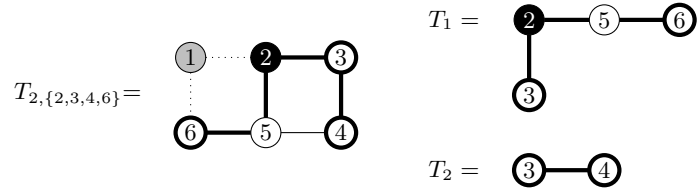

Now we fix column 2. Since $A_{22}=0$ after fixing the first column (first image in Figure \ref{fig:app:linUp2_0}) so we try to propagate 1 from a row below it. In this case the nearest node to $2$ in the graph, that has a 1 in the matrix, is 3. So we apply $\CNOT_{32}$ and the corresponding row operation is as follows:
\begin{eqnarray}
 \mathcal{Y}_1^{2'}&=&\{\CNOT_{32}\}  \nonumber \\
 \mathcal{A}_1^{2'}&=&\{A[2,.]\leftarrow A[2,.]\oplus A[3,.]\}    \nonumber 
\end{eqnarray}
The matrix $A$ after this operation is as shown in the beginning of Figure \ref{fig:app:linUp2}. Now the pivot is at 2 and the set of terminal nodes is $\mathcal{S}=\{2,3,4,6\}$. We draw the Steiner tree $T_{2,\mathcal{S}}$ in the graph $G\setminus\{1,\}$ and divide it into sub-trees as shown in Figure \ref{fig:app:linUp2}, using the same procedure as described before. Then we traverse the sub-trees using ROW-OP (Algorithm \ref{alg:rowOP}) and get the following sequence of CNOTs and the corresponding row operations for $A$. 
\begin{eqnarray}
 \mathcal{Y}_1^2&=&\{\CNOT_{34},\CNOT_{23},\CNOT_{25},\CNOT_{56},\CNOT_{25}\}   \nonumber \\
 \mathcal{A}_1^2&=&\{A[4,.]\leftarrow A[4,.]\oplus A[3,.], A[3,.]\leftarrow A[3,.]\oplus A[2,.], \nonumber \\
 && A[5,.]\leftarrow A[5,.]\oplus A[2,.], A[6,.]\leftarrow A[6,.]\oplus A[5,.], A[5,.]\leftarrow A[5,.]\oplus A[2,.] \}   \nonumber
\end{eqnarray}

\subsection*{Column 3}

\begin{figure}
\centering
\begin{subfigure}{0.3\textwidth}
\centering
 \footnotesize
$A= \begin{bmatrix}
     1 & 1 & \mathbf{0} & 1 & 1 & 0 \\
     0 & 1 & \mathbf{1} & 1 & 0 & 0 \\
     0 & 0 & \mathbf{1} & 1 & 0 & 1 \\
     0 & 0 & \mathbf{1} & 1 & 1 & 1 \\
     0 & 0 & \mathbf{1} & 0 & 0 & 0 \\
     0 & 0 & \mathbf{0} & 0 & 0 & 1
    \end{bmatrix}
$
\end{subfigure}
\hfill
\begin{subfigure}{0.65\textwidth}
 \centering
 \footnotesize
\begin{tikzpicture}
\node at (-0.5,0.5) {$T_{3,\{3,4,5\}}$=};
 \draw [fill=lightgray] (1,1) circle [radius=0.2];
  \node at (1,1) {$1$};
  \draw [dotted] (1.2,1)--(1.8,1); 
  \draw (1.2,0)--(1.8,0); 
  \draw [fill=lightgray] (2,1) circle [radius=0.2];
  \node at (2,1) {$2$};
  \draw [dotted] (2.2,1)--(2.8,1); 
  \draw [ultra thick] (2.2,0)--(2.8,0); 
  \draw [fill=black] (3,1) circle [radius=0.2];
  \node [white] at (3,1) {$3$};
  \draw (1,0) circle [radius=0.2];
  \node at (1,0) {$6$};
  \draw [ultra thick] (2,0) circle [radius=0.2];
  \node at (2,0) {$5$};
  \draw [ultra thick] (3,0) circle [radius=0.2];
  \node at (3,0) {$4$};
  \draw [dotted] (1,0.8)--(1,0.2);    
  \draw [ultra thick] (3,0.8)--(3,0.2);    
  \draw [dotted] (2,0.8)--(2,0.2); 
  
  \node at (4.5,1) {$T_1=$};
  \draw [fill=black] (5.5,1) circle [radius=0.2];
  \node [white] at (5.5,1) {$3$};
  \draw [ultra thick] (5.7,1)--(6.3,1);
  \draw [ultra thick] (6.5,1) circle [radius = 0.2];
  \node at (6.5,1) {$4$};
  
  \node at (4.5,0) {$T_2=$};
  \draw [ultra thick] (5.5,0) circle [radius=0.2];
  \node at (5.5,0) {$4$};
  \draw [ultra thick] (5.7,0)--(6.3,0);
  \draw [ultra thick] (6.5,0) circle [radius=0.2];
  \node at (6.5,0) {$5$};
  
\end{tikzpicture}
\end{subfigure}
\caption{The Steiner tree $T_{3,\mathcal{S}}$ with pivot at $3$ and terminals $\mathcal{S}=\{3,4,5\}$. $T_1$ and $T_2$ are the sub-trees built from it.}
\label{fig:app:linUp3}
\end{figure}
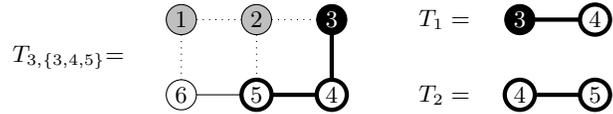

The state of $A$ before fixing column 3 i.e. after applying all the previous operations is as shown in the beginning of Figure \ref{fig:app:linUp3}. We draw the Steiner tree $T_{3,\mathcal{S}}$ in the graph $G\setminus\{1,2\}$ using the set of terminals $\mathcal{S}=\{3,4,5\}$. Then we sub-divide into sub-trees and get the following sequence of CNOTS and row operations, after traversing them according to ROW-OP (Algorithm \ref{alg:rowOP}).
\begin{eqnarray}
 \mathcal{Y}_1^3&=&\{\CNOT_{45},\CNOT_{34}\}    \nonumber \\
 \mathcal{A}_1^3&=&\{A[5,.]\leftarrow A[5,.]\oplus A[4,.],A[4,.]\leftarrow A[4,.]\oplus A[3,.]\}    \nonumber
\end{eqnarray}

\subsection*{Column 4}

\begin{figure}
\centering
\begin{subfigure}{0.3\textwidth}
\centering
 \footnotesize

$A= \begin{bmatrix}
     1 & 1 & 0 & \mathbf{1} & 1 & 0 \\
     0 & 1 & 1 & \mathbf{1} & 0 & 0 \\
     0 & 0 & 1 & \mathbf{1} & 0 & 1 \\
     0 & 0 & 0 & \mathbf{0} & 1 & 0 \\
     0 & 0 & 0 & \mathbf{1} & 1 & 1 \\
     0 & 0 & 0 & \mathbf{0} & 0 & 1
    \end{bmatrix}
$
\end{subfigure}
\hfill 
\begin{subfigure}{0.65\textwidth}
 \centering
 \footnotesize
\begin{tikzpicture}

 \draw [fill=lightgray] (1,1) circle [radius=0.2];
  \node at (1,1) {$1$};
  \draw [dotted] (1.2,1)--(1.8,1); 
  \draw (1.2,0)--(1.8,0); 
  \draw [fill=lightgray] (2,1) circle [radius=0.2];
  \node at (2,1) {$2$};
  \draw [dotted] (2.2,1)--(2.8,1); 
  \draw [ultra thick] (2.2,0)--(2.8,0); 
  \draw [fill=lightgray] (3,1) circle [radius=0.2];
  \node at (3,1) {$3$};
  \draw (1,0) circle [radius=0.2];
  \node at (1,0) {$6$};
  \draw [ultra thick] (2,0) circle [radius=0.2];
  \node at (2,0) {$5$};
  \draw [fill=black] (3,0) circle [radius=0.2];
  \node [white] at (3,0) {$4$};
  \draw [dotted] (1,0.8)--(1,0.2);    
  \draw [dotted] (3,0.8)--(3,0.2);    
  \draw [dotted] (2,0.8)--(2,0.2); 
  
\end{tikzpicture}
\end{subfigure}
\caption{Since $A_{44}=0$ so we apply $\CNOT_{54}$ and ``propagate'' $1$ from $A_{54}$ to $A_{44}$.}
\label{fig:app:linUp4_0}
\end{figure}
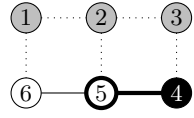

The matrix obtained after applying the previous operations in shown in Figure \ref{fig:app:linUp4_0}. Since $A_{44}=0$ we have to propagate a 1 from a row below it. In this case node 5 is the nearest to node 4 and has a 1. So we apply the following CNOT and the corresponding row operation.
\begin{eqnarray}
 \mathcal{Y}_1^{4'}&=&\{\CNOT_{54}\}    \nonumber \\
 \mathcal{A}_1^{4'}&=&\{A[4,.]\leftarrow A[4,.]\oplus A[5,.]\}  \nonumber
\end{eqnarray}
After this we extract a Steiner tree $T_{4,\mathcal{S}}$ from $G\setminus\{1,2,3\}$ with pivot 4 and set of terminals as $\mathcal{S}=\{4,5\}$. Traversing this tree (no sub-trees here) we get the following operations.
\begin{eqnarray}
 \mathcal{Y}_1^4&=&\{\CNOT_{45}\}   \nonumber \\
 \mathcal{A}_1^4&=&\{A[5,.]\leftarrow A[5,.]\oplus A[4,.]\} \nonumber
\end{eqnarray}
After these operations we find that column 5 and 6 are fixed. So $A$ has been reduced to an upper triangular form.

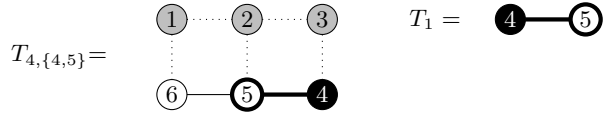
\begin{figure}
\centering
\begin{subfigure}{0.3\textwidth}
\centering
 \footnotesize
$A= \begin{bmatrix}
     1 & 1 & 0 & \mathbf{1} & 1 & 0 \\
     0 & 1 & 1 & \mathbf{1} & 0 & 0 \\
     0 & 0 & 1 & \mathbf{1} & 0 & 1 \\
     0 & 0 & 0 & \mathbf{1} & 0 & 1 \\
     0 & 0 & 0 & \mathbf{1} & 1 & 1 \\
     0 & 0 & 0 & \mathbf{0} & 0 & 1
    \end{bmatrix}
$
\end{subfigure}
\hfil
\begin{subfigure}{0.65\textwidth}
 \centering
 \footnotesize
\begin{tikzpicture}

\node at (-0.5,0.5) {$T_{4,\{4,5\}}$=};
 \draw [fill=lightgray] (1,1) circle [radius=0.2];
  \node at (1,1) {$1$};
  \draw [dotted] (1.2,1)--(1.8,1); 
  \draw (1.2,0)--(1.8,0); 
  \draw [fill=lightgray] (2,1) circle [radius=0.2];
  \node at (2,1) {$2$};
  \draw [dotted] (2.2,1)--(2.8,1); 
  \draw [ultra thick] (2.2,0)--(2.8,0); 
  \draw [fill=lightgray] (3,1) circle [radius=0.2];
  \node at (3,1) {$3$};
  \draw (1,0) circle [radius=0.2];
  \node at (1,0) {$6$};
  \draw [ultra thick] (2,0) circle [radius=0.2];
  \node at (2,0) {$5$};
  \draw [fill=black] (3,0) circle [radius=0.2];
  \node [white] at (3,0) {$4$};
  \draw [dotted] (1,0.8)--(1,0.2);    
  \draw [dotted] (3,0.8)--(3,0.2);    
  \draw [dotted] (2,0.8)--(2,0.2); 
  
    \node at (4.5,1) {$T_1=$};
  \draw [fill=black] (5.5,1) circle [radius=0.2];
  \node [white] at (5.5,1) {$4$};
  \draw [ultra thick] (5.7,1)--(6.3,1);
  \draw [ultra thick] (6.5,1) circle [radius = 0.2];
  \node at (6.5,1) {$5$};
  
\end{tikzpicture}
\end{subfigure}
\caption{The Steiner tree $T_{4,\mathcal{S}}$ with pivot at $4$ and terminals $\mathcal{S}=\{4,5\}$.}
\label{fig:app:linUp4}
\end{figure}

\subsection{Transpose and reducing to identity}

Now we transpose $A$, as shown in Figure \ref{fig:app:linDown1} and fix each column such that 1 remains only in the diagonal. 

\subsection*{Column 1}

We build a Steiner tree $T_{1,\{\mathcal{S}\}}$ (Figure \ref{fig:app:linDown1}) with pivot 1 and set of terminals $\mathcal{S}=\{1,2,4,5\}$. We divide into sub-trees according to Algorithm \ref{alg:separate} (SEPARATE) and do the traversals in each sub-tree, starting from the last one built, according to ROW-OP (Algorithm \ref{alg:rowOP}). This time we apply Bottom-Up-1, Top-Down-1, Bottom-Up-2 and Top-Down-2 and get the following sequence of CNOTs and row operations on $A$.
\begin{eqnarray}
 \mathcal{Y}_2^1&=&\{\CNOT_{54},\CNOT_{25},\CNOT_{12}\} \nonumber \\
 \mathcal{A}_2^1&=&\{A[4,.]\leftarrow A[4,.]\oplus A[5,.],A[5,.]\leftarrow A[5,.]\oplus A[2,.],A[2,.]\leftarrow A[2,.]\oplus A[1,.]\}   \nonumber
\end{eqnarray}
The purpose of applying all the sub-procedures is as follows. Let the parity at root and leaf before traversal is $x_r$ and $x_{\ell}$ respectively. After applying Bottom-Up-1, Top-Down-1, Bottom-Up-2, Top-Down-2 the parity at leaf is $x_{\ell}'=x_r\oplus x_{\ell}$. The parities at the other nodes remain unchanged.

Thus after Bottom-Up-1+Top-Down-1+Bottom-Up-2+Top-Down-2 on sub-tree $T_3$, which in this case is a single $\CNOT_{54}$, the parity at node 4 is $x_4'=x_4\oplus x_5$, and parity at 5 is unchanged. Similarly after applying the traversals in $T_2$, the parity at $5$ is $x_5'=x_5 \oplus x_2$ and the parity at 2 is unchanged. After traversing $T_1$ the parity at 2 is $x_2'=x_2\oplus x_1$ and the parity at 1 is unchanged. 

Now to avoid disturbing the upper-triangular 0s, we want that a node (row) should be XORed with a row with a higher index, which is clearly violated in case of $T_3$ where we have $x_4\oplus x_5$ at node 4. So we apply a correction procedure, by Algorithm \ref{alg:linTfSynth} (LINEAR-TF-SYNTH), after traversing all sub-trees. We take the shortest path from 5 to 4, which in this case is $T_3$, and again apply the same traversals. Thus we get the following sequence of CNOTs.
\begin{eqnarray}
 \mathcal{Y}_2^{1'}&=&\{\CNOT_{54}\}    \nonumber \\
 \mathcal{A}_2^{1'}&=&\{A[4,.]\leftarrow A[4,.]\oplus A[5,.]\}  \nonumber
\end{eqnarray}
Then the parity at 4 becomes $x_4' \oplus x_5' = (x_4\oplus x_5)\oplus (x_5\oplus x_2)=x_4\oplus x_2$, which satisfies the desired condition.

\begin{figure}
\centering
\begin{subfigure}{0.3\textwidth}
\centering
 \footnotesize

$A= \begin{bmatrix}
     \mathbf{1} & 0 & 0 & 0 & 0 & 0 \\
     \mathbf{1} & 1 & 0 & 0 & 0 & 0 \\
     \mathbf{0} & 1 & 1 & 0 & 0 & 0 \\
     \mathbf{1} & 1 & 1 & 1 & 0 & 0 \\
     \mathbf{1} & 0 & 0 & 0 & 1 & 0 \\
     \mathbf{0} & 0 & 1 & 1 & 0 & 1
    \end{bmatrix}
$
\end{subfigure}
\hfill
\begin{subfigure}{0.65\textwidth}
 \centering
 \footnotesize
\begin{tikzpicture}

\node at (-0.5,0.5) {$T_{1,\{1,2,4,5\}}$=};
 \draw [fill=black] (1,1) circle [radius=0.2];
  \node [white] at (1,1) {$1$};
  \draw [ultra thick] (1.2,1)--(1.8,1); 
  \draw (1.2,0)--(1.8,0); 
  \draw [ultra thick] (2,1) circle [radius=0.2];
  \node at (2,1) {$2$};
  \draw (2.2,1)--(2.8,1); 
  \draw [ultra thick] (2.2,0)--(2.8,0); 
  \draw (3,1) circle [radius=0.2];
  \node at (3,1) {$3$};
  \draw (1,0) circle [radius=0.2];
  \node at (1,0) {$6$};
  \draw [ultra thick] (2,0) circle [radius=0.2];
  \node at (2,0) {$5$};
  \draw [ultra thick] (3,0) circle [radius=0.2];
  \node at (3,0) {$4$};
  \draw (1,0.8)--(1,0.2);    
  \draw (3,0.8)--(3,0.2);    
  \draw [ultra thick] (2,0.8)--(2,0.2); 
  
  \node at (4.5,1.5) {$T_1=$};
  \draw [fill=black] (5.5,1.5) circle [radius=0.2];
  \node [white] at (5.5,1.5) {$1$};
  \draw [ultra thick] (5.7,1.5)--(6.3,1.5);
  \draw [ultra thick] (6.5,1.5) circle [radius = 0.2];
  \node at (6.5,1.5) {$2$};
  
  \node at (4.5,0.5) {$T_2=$};
  \draw [ultra thick] (5.5,0.5) circle [radius=0.2];
  \node at (5.5,0.5) {$2$};
  \draw [ultra thick] (5.7,0.5)--(6.3,0.5);
  \draw [ultra thick] (6.5,0.5) circle [radius=0.2];
  \node at (6.5,0.5) {$5$};
  
  \node at (4.5,-0.5) {$T_3=$};
  \draw [ultra thick] (5.5,-0.5) circle [radius=0.2];
  \node at (5.5,-0.5) {$5$};
  \draw [ultra thick] (5.7,-0.5)--(6.3,-0.5);
  \draw [ultra thick] (6.5,-0.5) circle [radius=0.2];
  \node at (6.5,-0.5) {$4$};  
\end{tikzpicture}
\end{subfigure}
\caption{The Steiner tree $T_{1,\mathcal{S}}$ with pivot at 1 and terminals $\mathcal{S}=\{1,2,4,5\}$. $T_1, T_2$ and $T_3$ are the sub-trees built from it.}
\label{fig:app:linDown1}
\end{figure}

\subsection*{Column 2}

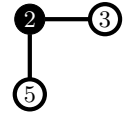
\begin{figure}
\centering
\begin{subfigure}{0.3\textwidth}
\centering
 \footnotesize
$A= \begin{bmatrix}
     1 & \mathbf{0} & 0 & 0 & 0 & 0 \\
     0 & \mathbf{1} & 0 & 0 & 0 & 0 \\
     0 & \mathbf{1} & 1 & 0 & 0 & 0 \\
     0 & \mathbf{0} & 1 & 1 & 0 & 0 \\
     0 & \mathbf{1} & 0 & 0 & 1 & 0 \\
     0 & \mathbf{0} & 1 & 1 & 0 & 1
    \end{bmatrix}
$
\end{subfigure}
\hfill
\begin{subfigure}{0.65\textwidth}
 \centering
 \footnotesize
\begin{tikzpicture}

\node at (-0.5,0.5) {$T_{2,\{2,3,5\}}$=};
 \draw [fill=lightgray] (1,1) circle [radius=0.2];
  \node at (1,1) {$1$};
  \draw [dotted] (1.2,1)--(1.8,1); 
  \draw (1.2,0)--(1.8,0); 
  \draw [fill=black] (2,1) circle [radius=0.2];
  \node [white] at (2,1) {$2$};
  \draw [ultra thick] (2.2,1)--(2.8,1); 
  \draw (2.2,0)--(2.8,0); 
  \draw [ultra thick] (3,1) circle [radius=0.2];
  \node at (3,1) {$3$};
  \draw (1,0) circle [radius=0.2];
  \node at (1,0) {$6$};
  \draw [ultra thick] (2,0) circle [radius=0.2];
  \node at (2,0) {$5$};
  \draw (3,0) circle [radius=0.2];
  \node at (3,0) {$4$};
  \draw [dotted] (1,0.8)--(1,0.2);    
  \draw (3,0.8)--(3,0.2);    
  \draw [ultra thick] (2,0.8)--(2,0.2); 
  
  \node at (4.5,1) {$T_1=$};
  \draw [fill=black] (5.5,1) circle [radius=0.2];
  \node [white] at (5.5,1) {$2$};
  \draw [ultra thick] (5.7,1)--(6.3,1);
  \draw [ultra thick] (6.5,1) circle [radius = 0.2];
  \node at (6.5,1) {$3$};
  \draw [ultra thick] (5.5,0) circle [radius=0.2];
  \node at (5.5,0) {$5$};
  \draw [ultra thick] (5.5,0.8)--(5.5,0.2);
\end{tikzpicture}
\end{subfigure}
\caption{The Steiner tree $T_{2,\mathcal{S}}$ with pivot at 2 and terminals $\mathcal{S}=\{2,3,5\}$.}
\label{fig:app:linDown2}
\end{figure}

The state of the matrix $A$ after applying the previous operations is shown in Figure \ref{fig:app:linDown2}. To fix column 2, we draw a Steiner tree $T_{2,\mathcal{S}}$ on the graph $G\setminus\{1\}$, with pivot 2 and set of terminals $\mathcal{S}=\{2,3,5\}$. We invoke procedure ROW-OP (Algorithm \ref{alg:rowOP}) and get the following sequence of CNOTs and row operations.
\begin{eqnarray}
 \mathcal{Y}_2^2&=&\{\CNOT_{23},\CNOT_{25}\}    \nonumber\\
 \mathcal{A}_2^2&=&\{A[3,.]\leftarrow A[3,.]\oplus A[2,.],A[5,.]\leftarrow A[5,.]\oplus A[2,.]\}    \nonumber
\end{eqnarray}
This time all the rows are XORed with a row with a higher index. So no correction procedures are required.

\subsection*{Column 3}

\begin{figure}
\centering
\begin{subfigure}{0.3\textwidth}
\centering
 \footnotesize
$A= \begin{bmatrix}
     1 & 1 & \mathbf{0} & 0 & 0 & 0 \\
     0 & 1 & \mathbf{0} & 0 & 0 & 0 \\
     0 & 0 & \mathbf{1} & 0 & 0 & 0 \\
     0 & 0 & \mathbf{1} & 1 & 0 & 0 \\
     0 & 0 & \mathbf{0} & 0 & 1 & 0 \\
     0 & 0 & \mathbf{1} & 1 & 0 & 1
    \end{bmatrix}
$
\end{subfigure}
\hfill
\begin{subfigure}{0.65\textwidth}
 \centering
 \footnotesize
\begin{tikzpicture}

\node at (-0.5,0.5) {$T_{3,\{3,4,6\}}$=};
 \draw [fill=lightgray] (1,1) circle [radius=0.2];
  \node at (1,1) {$1$};
  \draw [dotted] (1.2,1)--(1.8,1); 
  \draw [ultra thick] (1.2,0)--(1.8,0); 
  \draw [fill=lightgray] (2,1) circle [radius=0.2];
  \node at (2,1) {$2$};
  \draw [dotted] (2.2,1)--(2.8,1); 
  \draw [ultra thick] (2.2,0)--(2.8,0); 
  \draw [fill=black] (3,1) circle [radius=0.2];
  \node [white] at (3,1) {$3$};
  \draw [ultra thick] (1,0) circle [radius=0.2];
  \node at (1,0) {$6$};
  \draw (2,0) circle [radius=0.2];
  \node at (2,0) {$5$};
  \draw [ultra thick] (3,0) circle [radius=0.2];
  \node at (3,0) {$4$};
  \draw [dotted] (1,0.8)--(1,0);    
  \draw [ultra thick] (3,0.8)--(3,0.2);    
  \draw [dotted] (2,0.8)--(2,0.2); 
  
  \node at (4.5,1) {$T_1=$};
  \draw [fill=black] (5.5,1) circle [radius=0.2];
  \node [white] at (5.5,1) {$3$};
  \draw [ultra thick] (5.7,1)--(6.3,1);
  \draw [ultra thick] (6.5,1) circle [radius = 0.2];
  \node at (6.5,1) {$4$};
  
  \node at (4.5,0) {$T_2=$};
  \draw [ultra thick] (5.5,0) circle [radius=0.2];
  \node at (5.5,0) {$4$};
  \draw [ultra thick] (5.7,0)--(6.3,0);
  \draw (6.5,0) circle [radius=0.2];
  \node at (6.5,0) {$5$};
  \draw [ultra thick] (7.5,0) circle [radius=0.2];
  \node at (7.5,0) {$6$};
  \draw [ultra thick] (6.7,0)--(7.3,0);
  
\end{tikzpicture}
\end{subfigure}
\caption{The Steiner tree $T_{3,\mathcal{S}}$ with pivot at $3$ and terminals $\mathcal{S}=\{3,4,6\}$. $T_1$ and $T_2$ are the sub-trees built from it.}
\label{fig:app:linDown3}
\end{figure}
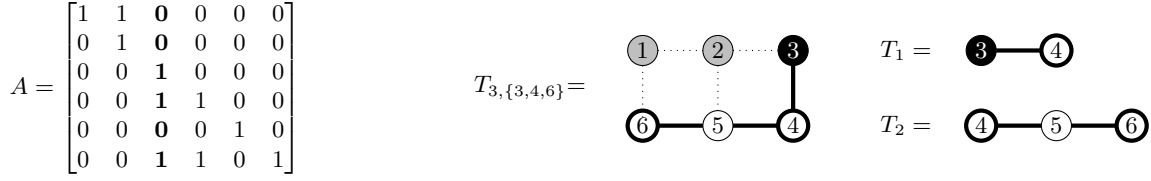

The state of $A$ after the previous operations is shown in Figure \ref{fig:app:linDown3}. As before we build a Steiner tree $T_{3,\mathcal{S}}$ on $G\setminus\{1,2\}$ with pivot 3 and set of terminals $\mathcal{S}=\{3,4,6\}$, which is further sub-divided into sub-trees $T_1$ and $T_2$. Applying the traversals according to ROW-OP (Algorithm \ref{alg:rowOP}) we get the following sequence of CNOTs and corresponding row operations.
\begin{eqnarray}
 \mathcal{Y}_2^3&=&\{\CNOT_{56},\CNOT_{45},\CNOT_{56},\CNOT_{45},\CNOT_{34}\}   \nonumber \\
 \mathcal{A}_2^3&=&\{A[6,.]\leftarrow A[6,.]\oplus A[5,.],A[5,.]\leftarrow A[5,.]\oplus A[4,.]  \nonumber \\
 && A[6,.]\leftarrow A[6,.]\oplus A[5,.], A[5,.]\leftarrow A[5,.]\oplus A[4,.], A[4,.]\leftarrow A[4,.]\oplus A[3,.] \} \nonumber
\end{eqnarray}
After all these operations we get $A=\id$.

\subsection{The complete circuit synthesizing a linear transformation matrix}

Let $\mathcal{Y}_1$ be the sequence (in order) of CNOTs obtained while reducing to upper triangular form. $\mathcal{Y}_2$ is the sequence of CNOTs obtained after transpose and before reducing to identity. $\mathcal{Y}_2'$ is the same sequence of CNOTs as in $\mathcal{Y}_2$, but with the control and target flipped. Then the final circuit is $\mathcal{Y}=\mathcal{Y}_2'\circ reverse(\mathcal{Y}_1)$, where $\circ$ means that we concatenate the reverse sequence in $\mathcal{Y}_1$ after $\mathcal{Y}_2'$. 

The number of CNOTs used by our algorithm is 26, while the algorithm in \cite{2020_NGM} used 43 CNOTs. So our algorithm fares much better in terms of CNOT requirement.

\section{An example for PHASE-NW-SYNTH}
\label{app:phase}

In this we illustrate Algorithm \ref{alg:phaseNwSynth} (PHASE-NW-SYNTH) with the connectivity graph $G$ given in Figure \ref{fig:app:lin0}. The input consists of the  following parity terms and their coefficients. The number of qubits $n=6$.
\begin{eqnarray}
 \mathcal{P}&=&\{(1,1\oplus x_1\oplus x_4\oplus x_5),(2,x_2\oplus x_3\oplus x_5\oplus x_6),(4,1\oplus x_4\oplus x_5\oplus x_6),(4,1\oplus x_1\oplus x_2\oplus x_6), \nonumber \\
 &&(6,1\oplus x_1\oplus x_2\oplus x_3),(7,1\oplus x_1\oplus x_2\oplus x_4\oplus x_6),(1,x_2\oplus x_4\oplus x_5)\}
\end{eqnarray}

\begin{figure}[h]
\centering
 \begin{subfigure}{0.3\textwidth}
  \centering
  \footnotesize
 $P= \begin{bmatrix}
     1 & 0 & 0 & 1 & 1 & 1 & 0 \\
     0 & 1 & 0 & 1 & 1 & 1 & 1 \\
     0 & 1 & 0 & 0 & 1 & 0 & 0 \\
     1 & 0 & 1 & 0 & 0 & 1 & 1 \\
     1 & 1 & 1 & 0 & 0 & 0 & 1 \\
     0 & 1 & 1 & 1 & 0 & 1 & 0 \\
     1 & 0 & 1 & 1 & 1 & 1 & 0 \\
     1 & 2 & 4 & 4 & 6 & 7 & 1
    \end{bmatrix}
$
\end{subfigure}
\hfill 
\begin{subfigure}{0.65\textwidth}
 \centering
 \footnotesize
\begin{tikzpicture}
\node at (3,0) {$G=$};
  \draw [thick] (4,1) circle [radius=0.25];
  \node at (4,1) {$1$};
  \draw [thick] (4.25,1)--(5.75,1); 
  \draw [thick] (4.25,-1)--(5.75,-1); 
  \draw [thick] (6,1) circle [radius=0.25];
  \node at (6,1) {$2$};
  \draw [thick] (6.25,1)--(7.75,1); 
  \draw [thick] (6.25,-1)--(7.75,-1); 
  \draw [thick] (8,1) circle [radius=0.25];
  \node at (8,1) {$3$};
  \draw [thick] (4,-1) circle [radius=0.25];
  \node at (4,-1) {$6$};
  \draw [thick] (6,-1) circle [radius=0.25];
  \node at (6,-1) {$5$};
  \draw [thick] (8,-1) circle [radius=0.25];
  \node at (8,-1) {$4$};
  \draw [thick] (4,0.75)--(4,-0.75);    
  \draw [thick] (8,0.75)--(8,-0.75);    
  \draw [thick] (6,0.75)--(6,-0.75); 
 \end{tikzpicture}
 \end{subfigure}
\caption{The parity matrix $P_{8\times 7}$ and connectivity graph $G$.}
\label{fig:app:ph0}
\end{figure}
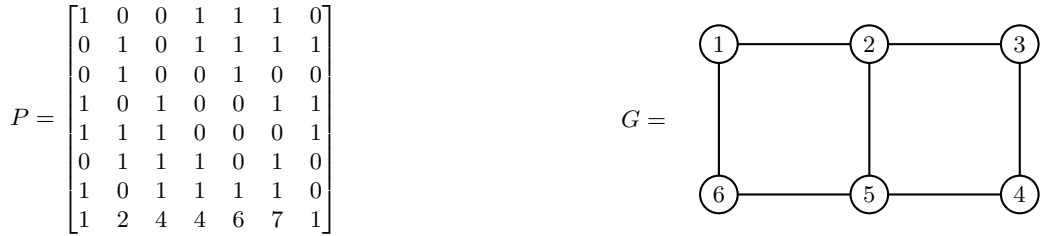

The parity matrix $P$ is a $8\times 7$ matrix where each column represents a parity term in $\mathcal{P}$. The first $6$ rows in each column encode the parity (without bit flip), the $7^{th}$ row encodes the bit flip term and the last row stores the coefficients. The matrix $P$ and the connecitvity graph $G$ has been shown in Figure \ref{fig:app:ph0}. We represent the set $B$ as a matrix where each column is a parity term ($p$) without the bit flip variable. We label column $i$ by ``$p_i$''. In this section superscripts denote the iteration we are in. 
$$B^{(0)}= \begin{bmatrix}
     \underline{p_1}&\underline{p_2}&\underline{p_3}&\underline{p_4}&\underline{p_5}&\underline{p_6}&\underline{p_7}    \\
     1 & 0 & 0 & 1 & 1 & 1 & 0 \\
     0 & 1 & 0 & 1 & 1 & 1 & 1 \\
     0 & 1 & 0 & 0 & 1 & 0 & 0 \\
     1 & 0 & 1 & 0 & 0 & 1 & 1 \\
     1 & 1 & 1 & 0 & 0 & 0 & 1 \\
     0 & 1 & 1 & 1 & 0 & 1 & 0 \\
    \end{bmatrix}
$$
We initialize an empty stack $\mathcal{K}$, $I=[6]$. We describe each iteration of the algorithm. A stack is a first-in and last-out data structure. From here on, we represent the stack $\mathcal{K}$ as an ordered set, where the rightmost element is the last one in or the element at the top of the stack. The leftmost element is the one at the bottom of the stack.

\subsection*{Iteration 1}

We select $j=2$ as the pivot row, according to step \ref{phaseNw:pivot} in Algorithm \ref{alg:phaseNwSynth}. The 0-cofactor ($B_0^1$) and 1-cofactor ($B_1^1$) are the columns which has $0$ and 1 in the $j^{the}$ row respectively. The stack $\mathcal{K}$ is as follows. 
\begin{eqnarray}
 B_1^1&=&\{p_2,p_4,p_5,p_6,p_7\} \quad \text{and}\quad B_0^1=\{p_1,p_3\}  \nonumber \\
 \mathcal{K}&=&\left\{(B_1^1,I\setminus\{2\},2),(B_0^1,I\setminus\{2\},\epsilon)\right\} \nonumber
\end{eqnarray}
The stack contains tuples $(B',I,i)$, in which the first element is a subset of columns of $B$. The second element $I\subseteq [6]$ indicates which rows in $B'$ have not been considered as ``pivot''. We divide a set of columns into 0 and 1-cofactor, depending upon the entries in the pivot row. The third element $i$ is the index of the pivot row, according to which the columns in $B'$ became part of a 1-cofactor for the first time. 

At the beginning, if no such pivot row is encountered for a set of columns (for example, in $0$-cofactors) then we initialize the third element to some $\epsilon\notin [6]$. Once third element has some value in $[6]$, it remains unchanged for the columns in $B'$ throughout the algorithm, even if $B'$ is further sub-divided.

\subsection*{Iteration 2}

We pop $(B_0^1,I\setminus\{2\},\epsilon)$ from the top of the stack $\mathcal{K}$. We select $j=1$ as the pivot row. The 0-cofactor, 1-cofactor and $\mathcal{K}$ are as follows.
\begin{eqnarray}
 B_1^2&=&\{p_1\} \quad \text{and} \quad B_0^2=\{p_3\}    \nonumber \\
 \mathcal{K}&=&\left\{(B_1^1,I\setminus\{2\},2),(B_1^2,I\setminus\{2,1\},1),(B_0^2,I\setminus\{2,1\},\epsilon)\right\}  \nonumber
\end{eqnarray}

\subsection*{Iteration 3}

We pop $(B_0^2,I\setminus\{2,1\},\epsilon)$ from the top of $\mathcal{K}$ and select $j=4$ as the pivot row. The 0-cofactor, 1-cofactor and $\mathcal{K}$ is as follows.
\begin{eqnarray}
 B_1^3&=&\{p_3\} \quad \text{and} \quad B_0^3=\emptyset   \nonumber   \\
 \mathcal{K}&=&\left\{(B_1^1,I\setminus\{2\},2),(B_1^2,I\setminus\{2,1\},1),(B_1^3,I\setminus\{2,1,4\},4)\right\}  \nonumber
\end{eqnarray}

\subsection*{Iteration 4}

\begin{figure}[h]
\centering
\begin{tikzpicture}
 \node at (1.5,0) {$T_{4,\{4,5,6\}}^{(4)}$=};
 \draw (3,1) circle [radius=0.25];
  \node at (3,1) {$1$};
  \draw (3.25,1)--(4.25,1); 
  \draw [ultra thick] (3.25,-0.5)--(4.25,-0.5); 
  \draw (4.5,1) circle [radius=0.25];
  \node at (4.5,1) {$2$};
  \draw (4.75,1)--(5.75,1); 
  \draw [ultra thick] (4.75,-0.5)--(5.75,-0.5); 
  \draw (6,1) circle [radius=0.25];
  \node at (6,1) {$3$};
  \draw [ultra thick] (3,-0.5) circle [radius=0.25];
  \node at (3,-0.5) {$6$};
  \draw [ultra thick] (4.5,-0.5) circle [radius=0.25];
  \node at (4.5,-0.5) {$5$};
  \draw [fill=black] (6,-0.5) circle [radius=0.25];
  \node [white] at (6,-0.5) {$4$};
  \draw (3,0.75)--(3,-0.25);    
  \draw (6,0.75)--(6,-0.25);    
  \draw (4.5,0.75)--(4.5,-0.25); 
  
    \node at (8,1) {$T_1^{flip}=$};
  \draw [ultra thick] (9,1) circle [radius=0.25];
  \node at (9,1) {$5$};
  \draw [ultra thick] (9.25,1)--(10.25,1);
  \draw [ultra thick] (10.5,1) circle [radius = 0.25];
  \node at (10.5,1) {$4$};
  
  \node at (8,0) {$T_2^{flip}=$};
  \draw [ultra thick] (9,0) circle [radius=0.25];
  \node at (9,0) {$6$};
  \draw [ultra thick] (9.25,0)--(10.25,0);
  \draw [ultra thick] (10.5,0) circle [radius = 0.25];
  \node at (10.5,0) {$5$};

\end{tikzpicture}
\caption{The Steiner tree $T_{4,\{4,5,6\}}^{(4)}$ (iteration 4). The sub-trees $T_1$ and $T_2$ are are flipped such that root becomes leaf and vice-versa. }
\label{fig:app:ph4}
\end{figure}

\begin{figure}
\centering 
\begin{subfigure}{0.4\textwidth}
\centering
\footnotesize
$B^{(4)}= \begin{bmatrix}
     \underline{p_1}&\underline{p_2}&\underline{p_3}&\underline{p_4}&\underline{p_5}&\underline{p_6}&\underline{p_7}    \\
     1 & 0 & 0 & 1 & 1 & 1 & 0 \\
     0 & 1 & 0 & 1 & 1 & 1 & 1 \\
     0 & 1 & 0 & 0 & 1 & 0 & 0 \\
     1 & 0 & 1 & 0 & 0 & 1 & 1 \\
     0 & 1 & 0 & 0 & 0 & 1 & 0 \\
     1 & 0 & 0 & 1 & 0 & 1 & 1 \\
    \end{bmatrix}
$
\caption{}
\end{subfigure}
\hfill 
\begin{subfigure}{0.5\textwidth}
\small
\centering
 \Qcircuit @C=1em @R=1em{
 \lstick{x_1} & \qw & \qw& \qw&\qw&\qw&\qw&\qw&\rstick{x_1} \\
 \lstick{x_2} & \qw & \qw&\qw&\qw&\qw&\qw&\qw&\rstick{x_2} \\
 \lstick{x_3} & \qw & \qw&\qw&\qw&\qw&\qw&\qw&\rstick{x_3} \\
 \lstick{x_4} & \qw & \qw& \qw&\targ&\qw&\gate{XZ}&\qw&\rstick{1\oplus x_4\oplus x_5\oplus x_6} \\
 \lstick{x_5} & \qw & \targ &\qw&\ctrl{-1}&\qw&\qw&\qw&\rstick{x_5\oplus x_6} \\
 \lstick{x_6} & \qw & \ctrl{-1} &\qw&\qw&\qw&\qw&\qw&\rstick{x_6} 
 }
\caption{}
 \end{subfigure}
  \caption{(a) $B$ after applying row operations in iteration 4. (b) The partial circuit obtained after applying the sequence of gates obtained from iteration 4. The variables on the left and right denote the parities before and after the application of the gates respectively.  }
 \label{fig:app:phCkt4}
\end{figure}

We pop $(B_1^3,I\setminus\{2,1,4\},4)$ from the top of $\mathcal{K}$. Since the third element of the tuple is an integer, so we go to step \ref{phaseNw:S'} in Algorithm \ref{alg:phaseNwSynth} and check if there is a set $\tset'$ of rows such that all columns in $B_1^3=\{p_3\}$ have a $1$. We find $\tset'=\{5,6\}$. We then build a Steiner tree $T_{4,\tset}^{(4)}$ on $G$ with set $\tset=\{4,5,6\}$ of terminals and ``pivot node'' as 4 (Figure \ref{fig:app:ph4}). Then we divide the tree into sub-trees as described in Algorithm \ref{alg:separate} (SEPARATE). If a sub-tree has multiple leaves we make each path as a separate sub-tree. We flip the root and leaf of each sub-tree and invoke Algorithm \ref{alg:rowOP} (ROW-OP) where we get the sequence of CNOTs according to the 4 traversals : Bottom-Up-1, Top-Down-1, Bottom-Up-2, Top-Down-2. The row operations invoked on $B$ depend only on the root and leaf of each sub-tree. 
\begin{eqnarray}
 \mathcal{Y}&=&\{\CNOT_{65},\CNOT_{54}\}    \nonumber \\
 \mathcal{A}&=&\{B[6,.]\leftarrow B[6,.]\oplus B[5,.],B[5,.]\leftarrow B[5,.]\oplus B[4,.]\}    \nonumber
\end{eqnarray}
$B$ after the row operations and the partial circuit has been shown in Figure \ref{fig:app:phCkt4}. We find that column $p_3$ has been fixed i.e. there is a single 1 and rest are 0. This implies that the $3^{rd}$ parity in $P$ is realized. Since $P_{73}=1$, so the bit-flip variable is set to $1$. The coefficient $P_{83}=4$. So we need to place $\X$ and $\Z$ gate at the point in the circuit where the parity has been realized. The stack $\mathcal{K}=\{(B_1^1,I\setminus\{2\},2),(B_1^2,I\setminus\{2,1\},1)\}$. Also, after this point we remove the column $p_3$ from $B$.

\subsection*{Iteration 5}

\begin{figure}[h]
\centering
\begin{tikzpicture}
 \node at (1.5,0) {$T_{1,\{1,4,6\}}^{(5)}$=};
 \draw [fill=black] (3,1) circle [radius=0.25];
  \node [white] at (3,1) {$1$};
  \draw (3.25,1)--(4.25,1); 
  \draw [ultra thick] (3.25,-0.5)--(4.25,-0.5); 
  \draw (4.5,1) circle [radius=0.25];
  \node at (4.5,1) {$2$};
  \draw (4.75,1)--(5.75,1); 
  \draw [ultra thick] (4.75,-0.5)--(5.75,-0.5); 
  \draw (6,1) circle [radius=0.25];
  \node at (6,1) {$3$};
  \draw [ultra thick] (3,-0.5) circle [radius=0.25];
  \node at (3,-0.5) {$6$};
  \draw (4.5,-0.5) circle [radius=0.25];
  \node at (4.5,-0.5) {$5$};
  \draw [ultra thick] (6,-0.5) circle [radius=0.25];
  \node at (6,-0.5) {$4$};
  \draw [ultra thick] (3,0.75)--(3,-0.25);    
  \draw (6,0.75)--(6,-0.25);    
  \draw (4.5,0.75)--(4.5,-0.25); 
  
    \node at (8,1) {$T_1^{flip}=$};
  \draw [ultra thick] (9,1) circle [radius=0.25];
  \node at (9,1) {$6$};
  \draw [ultra thick] (9.25,1)--(10.25,1);
  \draw [ultra thick] (10.5,1) circle [radius = 0.25];
  \node at (10.5,1) {$1$};
  
  \node at (8,0) {$T_2^{flip}=$};
  \draw [ultra thick] (9,0) circle [radius=0.25];
  \node at (9,0) {$4$};
  \draw [ultra thick] (9.25,0)--(10.25,0);
  \draw (10.5,0) circle [radius = 0.25];
  \node at (10.5,0) {$5$};
  \draw [ultra thick] (12,0) circle [radius = 0.25];
  \node at (12,0) {$6$};
  \draw [ultra thick] (10.75,0)--(11.75,0);
\end{tikzpicture}
\caption{The Steiner tree $T_{1,\{1,4,6\}}^{(5)}$ (iteration 5). Sub-trees $T_1$ and $T_2$ are flipped such that root becomes leaf and vice-versa. }
\label{fig:app:ph5}
\end{figure}

\begin{figure}
\centering 
\begin{subfigure}{0.45\textwidth}
\scriptsize
\centering
$B^{(5)}= \begin{bmatrix}
     \underline{p_1}&\underline{p_2}&\underline{p_4}&\underline{p_5}&\underline{p_6}&\underline{p_7}    \\
     1 & 0 & 1 & 1 & 1 & 0 \\
     0 & 1 & 1 & 1 & 1 & 1 \\
     0 & 1 & 0 & 1 & 0 & 0 \\
     0 & 0 & 1 & 0 & 0 & 0 \\
     0 & 1 & 0 & 0 & 1 & 0 \\
     0 & 0 & 0 & 1 & 0 & 1 \\
    \end{bmatrix}
$
\caption{}
\end{subfigure}
\hfill 
\begin{subfigure}{0.45\textwidth}
\centering
\scriptsize
 \Qcircuit @C=1em @R=1em{
 \lstick{x_1} &\qw&\qw&\qw&\qw&\targ&\gate{XT}&\qw&\rstick{1\oplus x_1\oplus x_4\oplus x_5} \\
 \lstick{x_2} &\qw&\qw&\qw&\qw&\qw&\qw&\qw&\rstick{x_2} \\
 \lstick{x_3} &\qw&\qw&\qw&\qw&\qw&\qw&\qw&\rstick{x_3} \\
 \lstick{1\oplus x_4\oplus x_5\oplus x_6}&\qw&\ctrl{1}&\qw&\ctrl{1}&\qw& \qw&\qw&\rstick{1\oplus x_4\oplus x_5\oplus x_6} \\
 \lstick{x_5\oplus x_6} &\ctrl{1}&\targ&\ctrl{1}&\targ&\qw&\qw&\qw&\rstick{x_5\oplus x_6} \\
 \lstick{x_6} &\targ&\qw&\targ&\qw&\ctrl{-5}&\qw&\qw&\rstick{x_4\oplus x_5} 
 }
\caption{}
 \end{subfigure}
  \caption{(a) $B$ after applying row operations in iteration 5. (b) The partial circuit obtained after applying the sequence of gates obtained from iteration 5. The variables on the left and right denote the parities before and after the application of the gates respectively.  }
 \label{fig:app:phCkt5}
\end{figure}
\begin{figure}[h]
\centering
\begin{tikzpicture}
 \node at (1.5,0) {$T_{2,\{2,6\}}^{(8)}$=};
 \draw (3,1) circle [radius=0.25];
  \node at (3,1) {$1$};
  \draw (3.25,1)--(4.25,1); 
  \draw [ultra thick] (3.25,-0.5)--(4.25,-0.5); 
  \draw [fill=black] (4.5,1) circle [radius=0.25];
  \node [white] at (4.5,1) {$2$};
  \draw (4.75,1)--(5.75,1); 
  \draw (4.75,-0.5)--(5.75,-0.5); 
  \draw (6,1) circle [radius=0.25];
  \node at (6,1) {$3$};
  \draw [ultra thick] (3,-0.5) circle [radius=0.25];
  \node at (3,-0.5) {$6$};
  \draw (4.5,-0.5) circle [radius=0.25];
  \node at (4.5,-0.5) {$5$};
  \draw (6,-0.5) circle [radius=0.25];
  \node at (6,-0.5) {$4$};
  \draw (3,0.75)--(3,-0.25);    
  \draw (6,0.75)--(6,-0.25);    
  \draw [ultra thick] (4.5,0.75)--(4.5,-0.25); 
    
  \node at (8,0) {$T_1^{flip}=$};
  \draw [ultra thick] (9,0) circle [radius=0.25];
  \node at (9,0) {$6$};
  \draw [ultra thick] (9.25,0)--(10.25,0);
  \draw (10.5,0) circle [radius = 0.25];
  \node at (10.5,0) {$5$};
  \draw [ultra thick] (12,0) circle [radius = 0.25];
  \node at (12,0) {$2$};
  \draw [ultra thick] (10.75,0)--(11.75,0);
\end{tikzpicture}
\caption{The Steiner tree $T_{2,\{2,6\}}^{(8)}$ (iteration 8). The single sub-tree is flipped such that root becomes leaf and vice-versa. }
\label{fig:app:ph8}
\end{figure}
We pop $(B_1^2,I\setminus\{2,1\},1)$ from $\mathcal{K}$ where $B_1^2=\{p_1\}$. As explained in the previous iteration we go to step \ref{phaseNw:S'} in Algorithm \ref{alg:phaseNwSynth} and find $\tset'=\{4,6\}$. Thus we build a Steiner tree $T_{1,\tset}^{(5)}$ on $G$ with set of terminals $\tset=\{1,4,6\}$ and pivot node at 1 (Figure \ref{fig:app:ph5}). Then we separate into two sub-trees and flip them, obtaining $T_1^{flip}$ and $T_2^{flip}$. After applying the traversals in Algorithm \ref{alg:rowOP} (ROW-OP) we get the following sequence of CNOTs and row operations.
\begin{eqnarray}
 \mathcal{Y}&=&\{\CNOT_{56},\CNOT_{45},\CNOT_{56},\CNOT_{45},\CNOT_{61}\}   \nonumber   \\
 \mathcal{A}&=&\left\{B[4,.]\leftarrow B[4,.]\oplus B[6,.], B[6,.]\leftarrow B[6,.]\oplus B[1,.]\right\}    \nonumber
\end{eqnarray}
Applying these operations we find that $p_1$ has been fixed (Figure \ref{fig:app:phCkt5}). So we remove this column, append the sequence of CNOTs and according to the bit flip variable value ($P_{71}=1$) and coefficient ($P_{81}=1$) we place $\X$ and $\T$ gate at the place where the $1^{st}$ parity in $P$ has been realized. Now stack has $\mathcal{K}=\{(B_1^1,I\setminus\{2,\},2)\}$.

\subsection*{Iteration 6}

We pop $(B_1^1,I\setminus\{2,\},2)$ from the top of $\mathcal{K}$, where $B_1^1=\{p_2,p_4,p_5,p_6,p_7\}$. We select $j=1$ as the pivot row according to step \ref{phaseNw:pivot} of Algorithm \ref{alg:phaseNwSynth} (The row which satisfies condition is $j=4$. Then the algorithm will terminate much earlier with lesser CNOTs. Initially, this was an error on our part, but later we realized that we can show some non-trivial steps with $j=1$, which will hopefully clarify some aspects. So we decided to stick to $j=1$.)
. Now the 0-cofactor, 1-cofactor and stack are as follows.
\begin{eqnarray}
 B_1^6&=&\{p_4,p_5,p_6\}\quad\text{and}\quad B_0^6=\{p_2,p_7\}    \nonumber \\
 \mathcal{K}&=&\left\{(B_1^6,I\setminus\{2,1\},2),(B_0^6,I\setminus\{2,1\},2)\right\}   \nonumber
\end{eqnarray}

\subsection*{Iteration 7}

We pop $(B_0^6,I\setminus\{2,1\},2)$ from the top of $\mathcal{K}$ and set $j=3$ as the pivot row. The 0-cofactor, 1-cofactor and $\mathcal{K}$ are as follows.
\begin{eqnarray}
 B_1^7&=&\{p_2\}\quad\text{and}\quad B_0^7=\{p_7\}  \nonumber \\
 \mathcal{K}&=&\left\{(B_1^6,I\setminus\{2,1\},2),(B_1^7,I\setminus\{2,1,3\},2),(B_0^7,I\setminus\{2,1,3\},2)\right\}  \nonumber
\end{eqnarray}

\subsection*{Iteration 8}

\begin{figure}
\centering 
\begin{subfigure}{0.45\textwidth}
\centering
\scriptsize
$B^{(8)}= \begin{bmatrix}
     \underline{p_2}&\underline{p_4}&\underline{p_5}&\underline{p_6}&\underline{p_7}    \\
     0 & 1 & 1 & 1 & 0 \\
     1 & 1 & 1 & 1 & 1 \\
     1 & 0 & 1 & 0 & 0 \\
     0 & 1 & 0 & 0 & 0 \\
     1 & 0 & 0 & 1 & 0 \\
     1 & 1 & 0 & 1 & 0 \\
    \end{bmatrix}
$
\caption{}
\end{subfigure}
\hfill 
\begin{subfigure}{0.45\textwidth}
\centering
\scriptsize
 \Qcircuit @C=1em @R=1em{
 \lstick{1\oplus x_1\oplus x_4\oplus x_5} &\qw &\qw&\qw&\qw&\qw& \rstick{1\oplus x_1\oplus x_4\oplus x_5}\\
 \lstick{x_2} & \qw&\targ&\qw&\targ&\gate{T}&\rstick{x_2\oplus x_4\oplus x_5} \\
 \lstick{x_3} & \qw&\qw&\qw&\qw&\qw&\rstick{x_3} \\
 \lstick{1\oplus x_4\oplus x_5\oplus x_6} & \qw&\qw &\qw&\qw&\qw&\rstick{1\oplus x_4\oplus x_5\oplus x_6} \\
 \lstick{x_5\oplus x_6} & \qw&\ctrl{-3} &\targ&\ctrl{-3}&\targ&\rstick{x_5\oplus x_6} \\
 \lstick{x_4\oplus x_5} & \qw&\qw&\ctrl{-1}&\qw&\ctrl{-1}&\rstick{x_4\oplus x_5} 
 }
\caption{}
 \end{subfigure}
  \caption{(a) $B$ after applying row operations in iteration 8. (b) The partial circuit obtained after applying the sequence of gates obtained from iteration 8. The variables on the left and right denote the parities before and after the application of the gates respectively.  }
 \label{fig:app:phCkt8}
\end{figure}

We pop $(B_0^7,I\setminus\{2,1,3\},2)$ from $\mathcal{K}$. According to step \ref{phaseNw:S'}, we find $\tset'=\{6\}$. So we build a Steiner tree $T_{2,\tset}^{(8)}$ on $G$ (Figure \ref{fig:app:ph8}), with pivot at $2$ and set of terminals $\tset=\{2,6\}$. Following the traversals in Algorithm \ref{alg:rowOP} (ROW-OP) the sequence of CNOTs and row operations obtained as as follows.
\begin{eqnarray}
 \mathcal{Y}&=&\{\CNOT_{52},\CNOT_{65},\CNOT_{52},\CNOT_{65}\}  \nonumber \\
 \mathcal{A}&=&\{B[6,.]\leftarrow B[6,.]\oplus B[2,.]\} \nonumber
\end{eqnarray}
The state of $B$ and the partial circuit is shown in Fig \ref{fig:app:phCkt8}. According to the value of the bit-flip variable ($P_{77}=0$) and coefficient ($P_{87}=1$) we place a $\T$-gate at the place where the $7^{th}$ parity term of $P$ is realized. The stack is $\mathcal{K}=\{(B_1^6,I\setminus\{2,1\},2),(B_1^7,I\setminus\{2,1,3\},2)\}$.
Since $p_7$ is fixed, so we remove it from $B$.

\subsection*{Iteration 9}

\begin{figure}[h]
\centering
\begin{tikzpicture}
 \node at (1.5,0) {$T_{2,\{2,3,5,6\}}^{(9)}$=};
 \draw (3,1) circle [radius=0.25];
  \node at (3,1) {$1$};
  \draw (3.25,1)--(4.25,1); 
  \draw [ultra thick] (3.25,-0.5)--(4.25,-0.5); 
  \draw [fill=black] (4.5,1) circle [radius=0.25];
  \node [white] at (4.5,1) {$2$};
  \draw [ultra thick] (4.75,1)--(5.75,1); 
  \draw (4.75,-0.5)--(5.75,-0.5); 
  \draw [ultra thick] (6,1) circle [radius=0.25];
  \node at (6,1) {$3$};
  \draw [ultra thick] (3,-0.5) circle [radius=0.25];
  \node at (3,-0.5) {$6$};
  \draw [ultra thick] (4.5,-0.5) circle [radius=0.25];
  \node at (4.5,-0.5) {$5$};
  \draw (6,-0.5) circle [radius=0.25];
  \node at (6,-0.5) {$4$};
  \draw (3,0.75)--(3,-0.25);    
  \draw (6,0.75)--(6,-0.25);    
  \draw [ultra thick] (4.5,0.75)--(4.5,-0.25); 
    
  \node at (8,1) {$T_1^{flip}=$};
  \draw [ultra thick] (9,1) circle [radius=0.25];
  \node at (9,1) {$3$};
  \draw [ultra thick] (9.25,1)--(10.25,1);
  \draw [ultra thick] (10.5,1) circle [radius = 0.25];
  \node at (10.5,1) {$2$};

 \node at (8,0.2) {$T_2^{flip}=$};
  \draw [ultra thick] (9,0.2) circle [radius=0.25];
  \node at (9,0.2) {$5$};
  \draw [ultra thick] (9.25,0.2)--(10.25,0.2);
  \draw [ultra thick] (10.5,0.2) circle [radius = 0.25];
  \node at (10.5,0.2) {$2$};
  
   \node at (8,-0.6) {$T_3^{flip}=$};
  \draw [ultra thick] (9,-0.6) circle [radius=0.25];
  \node at (9,-0.6) {$6$};
  \draw [ultra thick] (9.25,-0.6)--(10.25,-0.6);
  \draw [ultra thick] (10.5,-0.6) circle [radius = 0.25];
  \node at (10.5,-0.6) {$5$};
\end{tikzpicture}
\caption{The Steiner tree $T_{2,\{2,3,5,6\}}^{(9)}$ (iteration 9). The sub-trees are flipped such that root becomes leaf and vice-versa. }
\label{fig:app:ph9}
\end{figure}

\begin{figure}
\centering 
\begin{subfigure}{0.45\textwidth}
\centering
\scriptsize
$B^{(9)}= \begin{bmatrix}
     \underline{p_2}&\underline{p_4}&\underline{p_5}&\underline{p_6}    \\
     0 & 1 & 1 & 1  \\
     1 & 1 & 1 & 1  \\
     0 & 1 & 0 & 1 \\
     0 & 1 & 0 & 0 \\
     0 & 1 & 1 & 0 \\
     0 & 1 & 0 & 0 \\
    \end{bmatrix}
$
\caption{}
\end{subfigure}
\hfill 
\begin{subfigure}{0.45\textwidth}
\centering
\scriptsize
 \Qcircuit @C=1em @R=1em{
 \lstick{1\oplus x_1\oplus x_4\oplus x_5} &\qw &\qw&\qw&\qw& \rstick{1\oplus x_1\oplus x_4\oplus x_5}\\
 \lstick{x_2\oplus x_4\oplus x_5} &\qw&\targ&\targ &\gate{S}&\rstick{x_2\oplus x_3\oplus x_5\oplus x_6}\\
 \lstick{x_3} & \qw&\qw&\ctrl{-1}&\qw&\rstick{x_3} \\
 \lstick{1\oplus x_4\oplus x_5\oplus x_6} &\qw& \qw&\qw&\qw&\rstick{1\oplus x_4\oplus x_5\oplus x_6} \\
 \lstick{x_5\oplus x_6} & \targ&\ctrl{-3}&\qw&\qw&\rstick{x_4\oplus x_6} \\
 \lstick{x_4\oplus x_5} &\ctrl{-1}&\qw &\qw&\qw&\rstick{x_4\oplus x_5} 
 }
\caption{}
 \end{subfigure}
  \caption{(a) $B$ after applying row operations in iteration 9. (b) The partial circuit obtained after applying the sequence of gates obtained from iteration 9. The variables on the left and right denote the parities before and after the application of the gates respectively.  }
 \label{fig:app:phCkt9}
\end{figure}

We pop $(B_1^7,I\setminus\{2,1,3\},2)$ from $\mathcal{K}$, where $B_1^7=\{p_2\}$. Similar to iteration 8, we build a Steiner tree $T_{2,\tset}$ on $G$ (Figure \ref{fig:app:ph9}) with pivot at $2$ and set of terminals $\tset=\{2,3,5,6\}$. We get the following sequence of CNOTs and row operations from ROW-OP(Algorithm \ref{alg:rowOP}).
\begin{eqnarray}
 \mathcal{Y}&=&\{\CNOT_{65},\CNOT_{52},\CNOT_{32}\}   \nonumber \\
 \mathcal{A}&=&\{B[6,.]\leftarrow B[6,.]\oplus B[5,.],B[5,.]\leftarrow B[5,.]\oplus B[2,.],B[3,.]\leftarrow B[3,.]\oplus B[2,.]\} \nonumber
\end{eqnarray}
The state of $B$ and the partial circuit has been shown in Figure \ref{fig:app:phCkt9}. We remove $p_2$ from $B$ since it is fixed. According to the bit flip variable ($P_{72}=0$) and coefficient ($P_{82}=2$) in $P$, we place $S$-gate at the place where the parity of column 2 in matrix $P$ is realized. The stack is
$\mathcal{K}=\{(B_1^6,I\setminus\{2,1\},2)\}$.

\subsection*{Iteration 10}

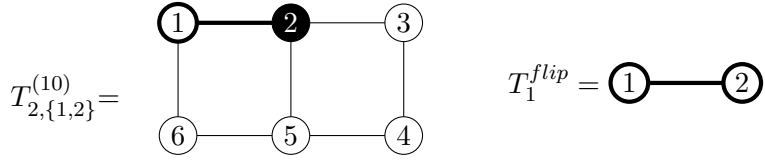
\begin{figure}[h]
\centering
\begin{tikzpicture}
 \node at (1.5,0) {$T_{2,\{1,2\}}^{(10)}$=};
 \draw [ultra thick] (3,1) circle [radius=0.25];
  \node at (3,1) {$1$};
  \draw [ultra thick] (3.25,1)--(4.25,1); 
  \draw (3.25,-0.5)--(4.25,-0.5); 
  \draw [fill=black] (4.5,1) circle [radius=0.25];
  \node [white] at (4.5,1) {$2$};
  \draw (4.75,1)--(5.75,1); 
  \draw (4.75,-0.5)--(5.75,-0.5); 
  \draw (6,1) circle [radius=0.25];
  \node at (6,1) {$3$};
  \draw (3,-0.5) circle [radius=0.25];
  \node at (3,-0.5) {$6$};
  \draw (4.5,-0.5) circle [radius=0.25];
  \node at (4.5,-0.5) {$5$};
  \draw (6,-0.5) circle [radius=0.25];
  \node at (6,-0.5) {$4$};
  \draw (3,0.75)--(3,-0.25);    
  \draw (6,0.75)--(6,-0.25);    
  \draw (4.5,0.75)--(4.5,-0.25); 

 \node at (8,0.2) {$T_1^{flip}=$};
  \draw [ultra thick] (9,0.2) circle [radius=0.25];
  \node at (9,0.2) {$1$};
  \draw [ultra thick] (9.25,0.2)--(10.25,0.2);
  \draw [ultra thick] (10.5,0.2) circle [radius = 0.25];
  \node at (10.5,0.2) {$2$};
\end{tikzpicture}
\caption{The Steiner tree $T_{2,\{1,2\}}^{(10)}$ (iteration 10). The single sub-tree is flipped such that root becomes leaf and vice-versa. }
\label{fig:app:ph10}
\end{figure}

\begin{figure}
\centering 
\begin{subfigure}{0.45\textwidth}
\centering
\scriptsize
$B^{(10)}= \begin{bmatrix}
     \underline{p_4}&\underline{p_5}&\underline{p_6}    \\
     0 & 0 & 0  \\
     1 & 1 & 1  \\
     1 & 0 & 1 \\
     1 & 0 & 0 \\
     1 & 1 & 0 \\
     1 & 0 & 0 \\
    \end{bmatrix}
$
\caption{}
\end{subfigure}
\hfill 
\begin{subfigure}{0.45\textwidth}
\centering
\scriptsize
 \Qcircuit @C=1em @R=1em{
 \lstick{1\oplus x_1\oplus x_4\oplus x_5}&\qw&\ctrl{1}&\qw&\rstick{1\oplus x_1\oplus x_4\oplus x_5}\\
 \lstick{x_2\oplus x_3\oplus x_5\oplus x_6}&\qw&\targ&\qw&\rstick{x_1\oplus x_2\oplus x_3\oplus x_4\oplus x_6}\\
 \lstick{x_3} &\qw&\qw&\qw&\rstick{x_3}\\
 \lstick{1\oplus x_4\oplus x_5\oplus x_6}&\qw&\qw&\qw&\rstick{1\oplus x_4\oplus x_5\oplus x_6} \\
 \lstick{x_4\oplus x_6}&\qw&\qw&\qw&\rstick{x_4\oplus x_6} \\
 \lstick{x_4\oplus x_5} &\qw&\qw&\qw& \rstick{x_4\oplus x_5} 
 }
\caption{}
 \end{subfigure}
  \caption{(a) $B$ after applying row operations in iteration 10. (b) The partial circuit obtained after applying the CNOT obtained from iteration 10. The variables on the left and right denote the parities before and after the application of the CNOT  respectively.  }
 \label{fig:app:phCkt10}
\end{figure}

We now pop $(B_1^6,I\setminus\{2,1\},2)$, where $B_1^6=\{p_4,p_5,p_6\}$. At step \ref{phaseNw:S'} of Algorithm \ref{alg:phaseNwSynth} we find $\tset'=\{1\}$. Thus we build a Steiner tree $T_{2,\tset}^{(10)}$ (Figure \ref
{fig:app:ph10}) on $G$ with pivot at 2 and terminal $\tset=\{1,2\}$. As before we flip the only sub-tree and obtain the following sequence of gates and row operations.
\begin{eqnarray}
 \mathcal{Y}&=&\{\CNOT_{12}\} \nonumber \\
 \mathcal{A}&=&\{B[1,.]\leftarrow B[1,.]\oplus B[2,.]\} \nonumber
\end{eqnarray}
No column is fixed (Figure \ref{fig:app:phCkt10}), so we go to step \ref{phaseNw:pivot} and further divide $B_1^6$ into 0-cofactor and 1-cofactor, by pivoting at $j=3$, and push these into the stack $\mathcal{K}$.
\begin{eqnarray}
 B_1^{10}&=&\{p_4,p_6\}\quad\text{and}\quad B_0^{10}=\{p_5\}  \nonumber \\
 \mathcal{K}&=&\{(B_1^{10},I\setminus\{2,1,3\},2),(B_0^{10},I\setminus\{2,1,3\},2)\}    \nonumber
\end{eqnarray}

\subsection*{Iteration 11}

\begin{figure}[h]
\centering
\begin{tikzpicture}
 \node at (1.5,0) {$T_{2,\{2,5\}}^{(11)}$=};
 \draw (3,1) circle [radius=0.25];
  \node at (3,1) {$1$};
  \draw (3.25,1)--(4.25,1); 
  \draw (3.25,-0.5)--(4.25,-0.5); 
  \draw [fill=black] (4.5,1) circle [radius=0.25];
  \node [white] at (4.5,1) {$2$};
  \draw (4.75,1)--(5.75,1); 
  \draw (4.75,-0.5)--(5.75,-0.5); 
  \draw (6,1) circle [radius=0.25];
  \node at (6,1) {$3$};
  \draw (3,-0.5) circle [radius=0.25];
  \node at (3,-0.5) {$6$};
  \draw [ultra thick] (4.5,-0.5) circle [radius=0.25];
  \node at (4.5,-0.5) {$5$};
  \draw (6,-0.5) circle [radius=0.25];
  \node at (6,-0.5) {$4$};
  \draw (3,0.75)--(3,-0.25);    
  \draw (6,0.75)--(6,-0.25);    
  \draw [ultra thick] (4.5,0.75)--(4.5,-0.25); 

 \node at (8,0.2) {$T_1^{flip}=$};
  \draw [ultra thick] (9,0.2) circle [radius=0.25];
  \node at (9,0.2) {$5$};
  \draw [ultra thick] (9.25,0.2)--(10.25,0.2);
  \draw [ultra thick] (10.5,0.2) circle [radius = 0.25];
  \node at (10.5,0.2) {$2$};
\end{tikzpicture}
\caption{The Steiner tree $T_{2,\{2,5\}}^{(11)}$ (iteration 11). The single sub-tree is flipped such that root becomes leaf and vice-versa. }
\label{fig:app:ph11}
\end{figure}

\begin{figure}
\begin{subfigure}{0.45\textwidth}
\centering
\scriptsize
$\footnotesize{  B^{(11)}= \begin{bmatrix}
     \underline{p_4}&\underline{p_5}&\underline{p_6}    \\
     0 & 0 & 0  \\
     1 & 1 & 1  \\
     1 & 0 & 1 \\
     1 & 0 & 0 \\
     0 & 0 & 1 \\
     1 & 0 & 0 \\
    \end{bmatrix}}
$
\caption{}
\end{subfigure}
\hfill 
\begin{subfigure}{0.45\textwidth}
\centering
\scriptsize
 \Qcircuit @C=1em @R=1em{
 \lstick{1\oplus x_1\oplus x_4\oplus x_5}&\qw&\qw&\qw&\rstick{1\oplus x_1\oplus x_4\oplus x_5}\\
 \lstick{x_1\oplus x_2\oplus x_3\oplus x_4\oplus x_6}&\qw&\targ&\gate{XS^{\dagger}}&\rstick{1\oplus x_1\oplus x_2\oplus x_3}\\
 \lstick{x_3} &\qw&\qw&\qw&\rstick{x_3}\\
 \lstick{1\oplus x_4\oplus x_5\oplus x_6}&\qw&\qw&\qw&\rstick{1\oplus x_4\oplus x_5\oplus x_6} \\
 \lstick{x_4\oplus x_6}&\qw&\ctrl{-3}&\qw&\rstick{x_4\oplus x_6} \\
 \lstick{x_4\oplus x_5} &\qw&\qw&\qw& \rstick{x_4\oplus x_5} 
 }
\caption{}
 \end{subfigure}
  \caption{(a) $B$ after applying row operations in iteration 11. (b) The partial circuit obtained after applying the gates obtained from iteration 11. The variables on the left and right denote the parities before and after the application of the gates respectively.  }
 \label{fig:app:phCkt11}
\end{figure}

We pop $(B_0^{10},I\setminus\{2,1,3\},2)$ from $\mathcal{K}$, where $B_0^{10}=\{p_5\}$. From the fifth column we find $\tset'=\{5\}$. Thus we build a Steiner tree $T_{2,\tset}^{(11)}$ (Figure \ref{fig:app:ph11}) on $G$ with pivot at $2$ and set of terminals $\tset=\{2,5\}$. Flipping and traversing we get the following sequence of CNOTs and row operations.
\begin{eqnarray}
 \mathcal{Y}&=&\{\CNOT_{52}\}   \nonumber \\
 \mathcal{A}&=&\{B[5,.]\leftarrow B[5,.]\oplus B[2,.]\} \nonumber
\end{eqnarray}
Now the state of $B$ and the appended circuit is shown in Figure \ref{fig:app:phCkt11}, where we see that $p_5$ has been fixed. Since the value of the bit variable is set ($P_{75}=1$) and coefficient is $P_{85}=6$, so we place a $\X$ and $\phase^{\dagger}$ gate at the place where the $5^{th}$ parity in $P$ is realized. The stack is $\mathcal{K}=\{(B_1^{(10)},I\setminus\{2,1,3\},2)\}$.

\subsection*{Iteration 12}

\begin{figure}[h]
\centering
\begin{tikzpicture}
 \node at (1.5,0) {$T_{2,\{2,3\}}^{(12)}$=};
 \draw (3,1) circle [radius=0.25];
  \node at (3,1) {$1$};
  \draw (3.25,1)--(4.25,1); 
  \draw (3.25,-0.5)--(4.25,-0.5); 
  \draw [fill=black] (4.5,1) circle [radius=0.25];
  \node [white] at (4.5,1) {$2$};
  \draw [ultra thick] (4.75,1)--(5.75,1); 
  \draw (4.75,-0.5)--(5.75,-0.5); 
  \draw [ultra thick] (6,1) circle [radius=0.25];
  \node at (6,1) {$3$};
  \draw (3,-0.5) circle [radius=0.25];
  \node at (3,-0.5) {$6$};
  \draw (4.5,-0.5) circle [radius=0.25];
  \node at (4.5,-0.5) {$5$};
  \draw (6,-0.5) circle [radius=0.25];
  \node at (6,-0.5) {$4$};
  \draw (3,0.75)--(3,-0.25);    
  \draw (6,0.75)--(6,-0.25);    
  \draw (4.5,0.75)--(4.5,-0.25); 

 \node at (8,0.2) {$T_1^{flip}=$};
  \draw [ultra thick] (9,0.2) circle [radius=0.25];
  \node at (9,0.2) {$3$};
  \draw [ultra thick] (9.25,0.2)--(10.25,0.2);
  \draw [ultra thick] (10.5,0.2) circle [radius = 0.25];
  \node at (10.5,0.2) {$2$};
\end{tikzpicture}
\caption{The Steiner tree $T_{2,\{2,3\}}^{(12)}$ (iteration 12). The single sub-tree is flipped such that root becomes leaf and vice-versa. }
\label{fig:app:ph12}
\end{figure}

\begin{figure}
\begin{subfigure}{0.45\textwidth}
\centering
\scriptsize
$\footnotesize{  B^{(12)}= \begin{bmatrix}
     \underline{p_4}&\underline{p_6}    \\
     0 & 0  \\
     1 & 1  \\
     0 & 0 \\
     1 & 0 \\
     0 & 1 \\
     1 & 0 \\
    \end{bmatrix}}
$
\caption{}
\end{subfigure}
\hfill 
\begin{subfigure}{0.45\textwidth}
\centering
\scriptsize
 \Qcircuit @C=1em @R=1em{
 \lstick{1\oplus x_1\oplus x_4\oplus x_5}&\qw&\qw&\rstick{1\oplus x_1\oplus x_4\oplus x_5}\\
 \lstick{1\oplus x_1\oplus x_2\oplus x_3}&\qw&\targ&\rstick{1\oplus x_1\oplus x_2}\\
 \lstick{x_3} &\qw&\ctrl{-1}&\rstick{x_3}\\
 \lstick{1\oplus x_4\oplus x_5\oplus x_6}&\qw&\qw&\rstick{1\oplus x_4\oplus x_5\oplus x_6} \\
 \lstick{x_4\oplus x_6}&\qw&\qw&\rstick{x_4\oplus x_6} \\
 \lstick{x_4\oplus x_5} &\qw&\qw& \rstick{x_4\oplus x_5} 
 }
\caption{}
 \end{subfigure}
  \caption{(a) $B$ after applying row operations in iteration 12. (b) The partial circuit obtained after applying the gates obtained from iteration 12. The variables on the left and right denote the parities before and after the application of the gates respectively.  }
 \label{fig:app:phCkt12}
\end{figure}

We pop $(B_1^{10},I\setminus\{2,1,3\},2)$ from $\mathcal{K}$, where $B_1^{10}=\{p_4,p_6\}$. At step $\ref{phaseNw:S'}$ of Algorithm \ref{alg:phaseNwSynth} we find that $\tset'=\{3\}$. So we build a Steiner tree $T_{2,\tset}$ (Figure \ref{fig:app:ph12}) on $G$ with pivot at 2 and set of terminals as $\tset=\{2,3\}$. After flipping and traversing we obtain the following sequence of CNOT gates and row operations.
\begin{eqnarray}
 \mathcal{Y}&=&\{\CNOT_{32}\}   \nonumber   \\
 \mathcal{A}&=&\{B[3,.]\leftarrow B[3,.]\oplus B[2,.]\} \nonumber
\end{eqnarray}
The state of $B$ and the appended circuit has been shown in Figure \ref{fig:app:phCkt12}. At step \ref{phaseNw:pivot} of Algorithm \ref{alg:phaseNwSynth} we sub-divide $B_1^{10}$ by pivoting at $j=4$. We get the following 0-cofactor, 1-cofactor, which we push into the stack.
\begin{eqnarray}
 B_1^{12}&=&\{p_4\}\quad\text{and}\quad B_0^{12}=\{p_6\}    \nonumber \\
 \mathcal{K}&=&\{(B_1^{12},I\setminus\{2,1,3,4\},2),(B_0^{12},I\setminus\{2,1,3,4\},2)\}    \nonumber
\end{eqnarray}

\subsection*{Iteration 13}

\begin{figure}[h]
\centering
\begin{tikzpicture}
 \node at (1.5,0) {$T_{2,\{2,5\}}^{(13)}$=};
 \draw (3,1) circle [radius=0.25];
  \node at (3,1) {$1$};
  \draw (3.25,1)--(4.25,1); 
  \draw (3.25,-0.5)--(4.25,-0.5); 
  \draw [fill=black] (4.5,1) circle [radius=0.25];
  \node [white] at (4.5,1) {$2$};
  \draw (4.75,1)--(5.75,1); 
  \draw (4.75,-0.5)--(5.75,-0.5); 
  \draw (6,1) circle [radius=0.25];
  \node at (6,1) {$3$};
  \draw (3,-0.5) circle [radius=0.25];
  \node at (3,-0.5) {$6$};
  \draw [ultra thick] (4.5,-0.5) circle [radius=0.25];
  \node at (4.5,-0.5) {$5$};
  \draw (6,-0.5) circle [radius=0.25];
  \node at (6,-0.5) {$4$};
  \draw (3,0.75)--(3,-0.25);    
  \draw (6,0.75)--(6,-0.25);    
  \draw [ultra thick] (4.5,0.75)--(4.5,-0.25); 

 \node at (8,0.2) {$T_1^{flip}=$};
  \draw [ultra thick] (9,0.2) circle [radius=0.25];
  \node at (9,0.2) {$5$};
  \draw [ultra thick] (9.25,0.2)--(10.25,0.2);
  \draw [ultra thick] (10.5,0.2) circle [radius = 0.25];
  \node at (10.5,0.2) {$2$};
\end{tikzpicture}
\caption{The Steiner tree $T_{2,\{2,5\}}^{(13)}$ (iteration 13). The single sub-tree is flipped such that root becomes leaf and vice-versa. }
\label{fig:app:ph13}
\end{figure}

\begin{figure}
\begin{subfigure}{0.45\textwidth}
\centering
\scriptsize
$\footnotesize{  B^{(13)}= \begin{bmatrix}
     \underline{p_4}&\underline{p_6}    \\
     0 & 0  \\
     1 & 1  \\
     0 & 0 \\
     1 & 0 \\
     1 & 0 \\
     1 & 0 \\
    \end{bmatrix}}
$
\caption{}
\end{subfigure}
\hfill 
\begin{subfigure}{0.45\textwidth}
\centering
\scriptsize
 \Qcircuit @C=1em @R=1em{
 \lstick{1\oplus x_1\oplus x_4\oplus x_5}&\qw&\qw&\qw&\rstick{1\oplus x_1\oplus x_4\oplus x_5}\\
 \lstick{1\oplus x_1\oplus x_2\oplus x_3}&\qw&\targ&\gate{T^{\dagger}}&\rstick{1\oplus x_1\oplus x_2\oplus x_4\oplus x_6}\\
 \lstick{x_3} &\qw&\qw&\qw&\rstick{x_3}\\
 \lstick{1\oplus x_4\oplus x_5\oplus x_6}&\qw&\qw&\qw&\rstick{1\oplus x_4\oplus x_5\oplus x_6} \\
 \lstick{x_4\oplus x_6}&\qw&\ctrl{-3}&\qw&\rstick{x_4\oplus x_6} \\
 \lstick{x_4\oplus x_5} &\qw&\qw&\qw& \rstick{x_4\oplus x_5} 
 }
\caption{}
 \end{subfigure}
  \caption{(a) $B$ after applying row operations in iteration 13. (b) The partial circuit obtained after applying the gates obtained from iteration 13. The variables on the left and right denote the parities before and after the application of the gates respectively.  }
 \label{fig:app:phCkt13}
\end{figure}

We pop $(B_0^{12},I\setminus\{2,1,3,4\},2)$, where $B_0^{12}=\{p_6\}$. We get $\tset'=\{5\}$ at step \ref{phaseNw:S'} of Algorithm \ref{alg:phaseNwSynth}. We build a Steiner tree $T_{2,\tset}^{(13)}$ on $G$ (Figure \ref{fig:app:ph13}) with pivot at 2 and set of terminals $\tset=\{2,5\}$. Then flipping and traversing we get the following sequence of gates and row operations.
\begin{eqnarray}
 \mathcal{Y}&=&\{\CNOT_{52}\}   \nonumber \\
 \mathcal{A}&=&\{B[5,.]\leftarrow B[5,.]\oplus B[2,.]\} \nonumber
\end{eqnarray}
The updated $B$ and the appended circuit has been shown in Figure \ref{fig:app:phCkt13}. The column $p_6$ is fixed, so we remove it from $B$. Now $P_{76}=1$ and we see that the circuit has the bit flip variable set to $1$. The coefficient $P_{86}=7$, so we place $T^{\dagger}$ at the place where the $6^{th}$ parity in $P$ is realized. Now the stack is $\mathcal{K}=\{(B_1^{12},I\setminus\{2,1,3,4\},2)\}$.

\subsection*{Iteration 14}

\begin{figure}[h]
\centering
\begin{tikzpicture}
 \node at (1.5,0) {$T_{2,\{2,4,5,6\}}^{(14)}$=};
 \draw (3,1) circle [radius=0.25];
  \node at (3,1) {$1$};
  \draw (3.25,1)--(4.25,1); 
  \draw [ultra thick] (3.25,-0.5)--(4.25,-0.5); 
  \draw [fill=black] (4.5,1) circle [radius=0.25];
  \node [white] at (4.5,1) {$2$};
  \draw (4.75,1)--(5.75,1); 
  \draw [ultra thick] (4.75,-0.5)--(5.75,-0.5); 
  \draw (6,1) circle [radius=0.25];
  \node at (6,1) {$3$};
  \draw [ultra thick] (3,-0.5) circle [radius=0.25];
  \node at (3,-0.5) {$6$};
  \draw [ultra thick] (4.5,-0.5) circle [radius=0.25];
  \node at (4.5,-0.5) {$5$};
  \draw [ultra thick] (6,-0.5) circle [radius=0.25];
  \node at (6,-0.5) {$4$};
  \draw (3,0.75)--(3,-0.25);    
  \draw (6,0.75)--(6,-0.25);    
  \draw [ultra thick] (4.5,0.75)--(4.5,-0.25); 

   \node at (8,1) {$T_1^{flip}=$};
  \draw [ultra thick] (9,1) circle [radius=0.25];
  \node at (9,1) {$5$};
  \draw [ultra thick] (9.25,1)--(10.25,1);
  \draw [ultra thick] (10.5,1) circle [radius = 0.25];
  \node at (10.5,1) {$2$};
  
 \node at (8,0.2) {$T_2^{flip}=$};
  \draw [ultra thick] (9,0.2) circle [radius=0.25];
  \node at (9,0.2) {$4$};
  \draw [ultra thick] (9.25,0.2)--(10.25,0.2);
  \draw [ultra thick] (10.5,0.2) circle [radius = 0.25];
  \node at (10.5,0.2) {$5$};
  
   \node at (8,-0.6) {$T_3^{flip}=$};
  \draw [ultra thick] (9,-0.6) circle [radius=0.25];
  \node at (9,-0.6) {$6$};
  \draw [ultra thick] (9.25,-0.6)--(10.25,-0.6);
  \draw [ultra thick] (10.5,-0.6) circle [radius = 0.25];
  \node at (10.5,-0.6) {$5$};
\end{tikzpicture}
\caption{The Steiner tree $T_{2,\{2,4,5,6\}}^{(14)}$ and its sub-trees, flipped such that root becomes leaf and vice-versa (iteration 14). }
\label{fig:app:ph14}
\end{figure}

\begin{figure}
\begin{subfigure}{0.45\textwidth}
\centering
\scriptsize
$\footnotesize{  B^{(14)}= \begin{bmatrix}
     \underline{p_4}    \\
     0  \\
     1  \\
     0 \\
     0 \\
     0 \\
     0 \\
    \end{bmatrix}}
$
\caption{}
\end{subfigure}
\hfill 
\begin{subfigure}{0.45\textwidth}
\centering
\scriptsize
 \Qcircuit @C=1em @R=1em{
 \lstick{1\oplus x_1\oplus x_4\oplus x_5}&\qw&\qw&\qw&\qw&\rstick{1\oplus x_1\oplus x_4\oplus x_5}\\
 \lstick{1\oplus x_1\oplus x_2\oplus x_4\oplus x_6}&\qw&\qw&\targ&\gate{Z}&\rstick{1\oplus x_1\oplus x_2\oplus x_6}\\
 \lstick{x_3} &\qw&\qw&\qw&\qw&\rstick{x_3}\\
 \lstick{1\oplus x_4\oplus x_5\oplus x_6}&\qw&\ctrl{1}&\qw&\qw&\rstick{1\oplus x_4\oplus x_5\oplus x_6} \\
 \lstick{x_4\oplus x_6}&\targ&\targ&\ctrl{-3}&\qw&\rstick{x_4\oplus x_6} \\
 \lstick{x_4\oplus x_5} &\ctrl{-1}&\qw&\qw&\qw& \rstick{x_4\oplus x_5} 
 }
\caption{}
 \end{subfigure}
  \caption{(a) $B$ after applying row operations in iteration 14. (b) The partial circuit obtained after applying the gates obtained from iteration 14. The variables on the left and right denote the parities before and after the application of the gates respectively.  }
 \label{fig:app:phCkt14}
\end{figure}

We pop $(B_1^{12},I\setminus\{2,1,3,4\},2)$ from $\mathcal{K}$, where $B_1^{12}=\{p_4\}$. Now $\tset'=\{4,5,6\}$. So we build a Steiner tree $T_{2,\tset}^{(14)}$ on $G$ (Figure \ref{fig:app:ph14}) with pivot at 2 and set of terminals $\tset=\{2,4,5,6\}$. After flipping the 3 sub-trees and traversing according to ROW-OP (Algorithm \ref{alg:rowOP}) we get the following sequence of CNOTs and row operations.
\begin{eqnarray}
 \mathcal{Y}&=&\{\CNOT_{65},\CNOT_{45},\CNOT_{52}\} \nonumber \\
 \mathcal{A}&=&\{B[6,.]\leftarrow B[6,.]\oplus B[5,.],B[4,.]\leftarrow B[4,.]\oplus B[5,.],B[5,.]\leftarrow B[5,.]\oplus B[2,.]\} \nonumber
\end{eqnarray}
The state of $B$ and the appended circuit has been shown in Figure \ref{fig:app:phCkt14}. We find that the remaining column $p_4$ has been fixed. Depending upon the parity realized and the coefficient (obtained from $P$) we place a $\Z$ gate at the appropriate position in the circuit.

\end{document}